\documentclass[aps,pre,twocolumn,showpacs,nofootinbib,superscriptaddress]{revtex4-1}

\usepackage{amsmath}
\usepackage{amssymb}
\usepackage{graphicx}
  \usepackage{hyperref}
\usepackage[arrow, matrix, curve]{xy}
\usepackage{bm}
\usepackage{yhmath}
\usepackage{color}
\usepackage{url}
\newcommand\diff{\mathrm{d}}

\usepackage[usenames,dvipsnames]{xcolor}
\hypersetup{colorlinks=true, linkcolor=BrickRed, urlcolor=blue!50!black, citecolor=blue!50!black}

\usepackage[normalem]{ulem}

\makeatletter
\newcommand\hide@visible[1]{%
  \bgroup\fboxsep=.3ex\colorbox{Gray}{begin hide}%
  #1\colorbox{Gray}{end hide}\egroup%
}
\newcommand\hide@hidden[1]{%
  \bgroup\fboxsep=.3ex\colorbox{Gray}{hidden text}%
}
\newcommand\hide@invisible[1]{}
\newcommand\makevisible{\let\hide\hide@visible}
\newcommand\makehidden{\let\hide\hide@hidden}
\newcommand\makeinvisible{\let\hide\hide@invisible}
\makeatother
\makehidden

\usepackage{etoolbox}
\let\bbordermatrix\bordermatrix
\patchcmd{\bbordermatrix}{8.75}{4.75}{}{}
\patchcmd{\bbordermatrix}{\left(}{\left[}{}{}
\patchcmd{\bbordermatrix}{\right)}{\right]}{}{}

\begin{document}


\title{Glassy dynamics in confinement: Planar and bulk limit of the mode-coupling theory
}





\author{Simon Lang}
\affiliation{Institut f{\"u}r Theoretische Physik, Leopold-Franzens-Universit\"at Innsbruck,
Technikerstra{\ss}e 25/2, A-6020 Innsbruck, Austria}

\author{Rolf Schilling}
\affiliation{Institut f{\"u}r Physik, Johannes Gutenberg-Universit{\"a}t Mainz,
 Staudinger Weg 7, 55099 Mainz, Germany}

\author{Thomas Franosch}
\affiliation{Institut f{\"u}r Theoretische Physik, Leopold-Franzens-Universit\"at Innsbruck,
Technikerstra{\ss}e 25/2, A-6020 Innsbruck, Austria}
\email[]{thomas.franosch@uibk.ac.at}
\date{\today}

\begin{abstract}
We demonstrate how the matrix-valued mode-coupling theory of the glass transition and glassy dynamics in planar confinement converges to the corresponding theory for two-dimensional (2D) planar  and the three-dimensional bulk liquid,  provided the wall potential satisfies certain conditions.
Since the mode-coupling theory relies on the static properties as input, the emergence of a homogeneous limit for the  matrix-valued intermediate scattering functions is directly connected to the convergence of the corresponding static quantities to their conventional counterparts. We show that the 2D limit is more subtle than the bulk limit, in particular, the in-planar dynamics decouples from the  motion perpendicular to the walls. We investigate the frozen-in parts of the intermediate scattering function in the glass state and find that the limits time $t\to \infty$ and effective wall separation $L\to 0$ do not commute due to the mutual coupling of the residual transversal and lateral force kernels.
\end{abstract}





\maketitle

\section{Introduction}

Confined liquids have been studied extensively in physics, in particular their phase behavior~\cite{Lowen:2001,Schmidt:1996,*Schmidt:1997}, dynamical properties~\cite{Alba:2006,Klafter:Restricted,Mittal:2008,Nugent:2007,Fehr:1995,Krishnan:2012,Gallo:2000a,*Gallo:2000b,*Gallo:2009,*Gallo:2012,Edmond:2012,
Eral:2009,Ingebrigtsen:2013,Scheidler:2000b,*Scheidler:2000a,*Scheidler:2002}, and their structural characterization~\cite{Evans:1990,Antonchenko:1984,Nygard:2012,*Nygard:2013}. Furthermore confined liquids are intermediate between a low-dimensional and bulk liquid and display  an intriguing interplay of near-range local ordering and the confining length.
In case of a slit geometry, i.e., two parallel flat hard walls with effective separation $L$ one can elaborate the full range from two-dimensional (2D) up to three-dimensional (3D) liquid behavior by varying the wall separation  from zero to infinity. For instance, for equilibrium phase transitions one can study the crossover~\cite{Qi:2014} from the Kosterlitz-Thouless transition~\cite{Strandburg:1988,Kosterlitz:1973,Young:1979,Nelson:1979,Bernard:2011,Kapfer:2014} in 2D liquids to the conventional
phase transitions of a 3D liquid.
Similarly, the crossover behavior of the glassy dynamics of a quasi-two-dimensional confined liquid towards a bulk system provides insight in the nature of the mechanism of structural arrest.

In \emph{bulk} systems many features of the slowing down of structural relaxation   upon cooling or compression have been rationalized in terms of the mode-coupling theory of the glass transition (MCT) developed by G\"otze and co-workers~\cite{Bengtzelius:1984,Goetze:Complex_Dynamics}. The predictions of MCT include the emergence of a non-trivial long-time limit of the intermediate scattering functions, called glass-form factors  associated with two-time fractals in the vicinity of the critical point.  Particularly, MCT entails a sharp dynamical glass transition ,e.g., for a hard-sphere liquid at a critical packing fraction.
Although in nature this transition is smeared, the various predictions of MCT
in three dimensions have been confirmed in the past two decades by experiments and
computer simulations~\cite{Goetze:Complex_Dynamics,Goetze:1999}.

In order to study the dependence of the glass transition on the spatial dimension
MCT has  also been worked out for two-dimensional single-component~\cite{Bayer:2007} and binary liquids~\cite{Hajnal:2009,Hajnal:2011,Weysser:2011}. Qualitative~\cite{Bayer:2007,Hajnal:2011,Weysser:2011} and quantitative \cite{Weysser:2011} comparisons of the MCT results for these 2D liquids with those from experiments~\cite{Koenig:2005} and simulations~\cite{Weysser:2011} have been reported. The two-dimensional systems are found to behave qualitatively similar to their three-dimensional counterparts.

Motivated by the numerous experimental and simulational results for the glass transition of confined liquids~\cite{Alba:2006,Fehr:1995,Krishnan:2012,Gallo:2000a,*Gallo:2000b,*Gallo:2009,*Gallo:2012,Nugent:2007,Edmond:2012,Eral:2009,Ingebrigtsen:2013,Scheidler:2000b,*Scheidler:2000a,*Scheidler:2002}, mostly for a slit geometry,
MCT has  been generalized recently to describe dense liquids in planar confinement~\cite{Lang:2010,Lang:2012,Lang:2014b}. In contrast to the MCT for the glassy dynamics in bulk or disordered structures~\cite{Krakoviack:2005,*Krakoviack:2007,*Krakoviack:2009,*Krakoviack:2011,Szamel:2013}, the inhomogeneous packing in the slit requires to consider symmetry-adapted matrix-valued intermediate scattering functions to characterize the density fluctuations of the confined liquid. In the case of hard spheres, the transition line,  separating the regime of collective frozen-in states from liquid states, has been determined as a function of the slit width. An intriguing multiple reentrant transition for wall separations on the scale of a few particle diameters has been predicted and corroborated by recent molecular-dynamics simulations with controlled polydispersity~\cite{Mandal:2014}.

While MCT has been successfully tested for planar, bulk and confined liquids, the natural question arises: Does MCT for liquids in planar confinement for $L \to 0$ and $L \to \infty$ properly converge to MCT for 2D and 3D liquids, respectively? Particularly, one would like to know if the dynamical behavior of a strongly confined liquid is approximately described by a two-dimensional system.
To provide an answer is the major goal of the present work.
 In particular, we will demonstrate how the matrix-valued MCT due to the inhomogeneous structure of the confined liquid reduces to the MCT of scalar intermediate scattering functions for the two-dimensional~\cite{Bayer:2007} and the bulk case~\cite{Goetze:Complex_Dynamics}.
In contrast to the bulk limit, the planar limit $L \to 0$ turns out to be rather subtle.

The outline of this work is as follows.
In the next section we introduce the model, the quantities of interest, and  recall the
 equations of motion for confined liquids within the MCT approximation. In Sec.\ref{Sec:2DLimit}  we discuss the behavior of the MCT functional  for small wall separation in terms of the proper convergence of the static properties towards their two-dimensional counterparts demonstrated recently~\cite{Lang:2014a}. In particular, we show that the planar MCT is recovered for all finite times in the limit of vanishing plate separation. Next we study the fixed-point equation for the glass-form factors for small slit widths. In Sec.~\ref{Sec:3DLimit} we demonstrate that the MCT equations include the bulk behavior as limiting case as the wall separation becomes infinitely large. Section~\ref{Sec:Summary} provides a critical assessment of the different convergences and possible implications for the glassy dynamics in extreme confinement.

\section{Confined liquids: Basic quantities and MCT}
The microscopic setup, the derivations of the equations of motion of the relevant dynamic quantities of interest is detailed in Ref.~\cite{Lang:2012}; here  we summarize  its main features, in order to keep the present paper self-contained.

Consider a liquid of $N$ particles of mass $m$ between two parallel, planar walls with cross section $A$ and separation $H$.  Then a point in phase space  is specified by the set of coordinates parallel and perpendicular to the wall  $\vec{x}_n = ( \vec{r}_n, z_n)$ and corresponding momenta $\vec{p}_n = (\vec{P}_n, P_n^z)$, $n=1,\ldots, N$.  We use impenetrable walls with additional wall potential of  the form $U(\{z_{n}\};L)  =\sum_{n=1}^{N} {\cal U}(z_{n};L)$ and
\begin{align}
{\cal U}(z;L) =
\begin{cases}
  {\cal U}_{\text{w}}(z) & \text{for} |z| \leq L/2, \\
    \infty & \text{for} |z|> L/2.  \\
\end{cases}
\end{align}
Here,  the effective wall separation $L$ is introduced as the transverse length accessible to the particles, and therefore
we distinguish between point particles and hard spheres,
\begin{equation}
L=
\begin{cases}
    H-\sigma ,& \text{hard spheres},  \\
    H, & \text{point particles}.
\end{cases}
\end{equation}
 The interaction energy of a  particle with  either of the  walls
${\cal U}_{\text{w}}(z)$ is assumed to be smooth, with possible divergencies at $z\equiv \pm L/2$. In principle, both walls can interact differently with the liquid, i.e., asymmetric wall potentials ${\cal U}_{\text{w}}(z)={\cal U}_{-}(L/2+z)+{\cal U}_{+}(L/2-z) $ with ${\cal U}_{-}(x)\neq{\cal U}_{+}(x)$ are allowed. Periodic boundary conditions are imposed
parallel to the walls in $x$-$y$ direction. We suppress the parametric dependence on the wall separation $L$ in the following to allow for a compact notation.  The pair interactions ${\cal V}(\vec{x})\equiv {\cal V}(\vec{r},z)$ depend only  on the mutual distance $|\vec{x}|$, i.e., $V(\{ \vec{x}_{n} \})= \sum_{n < m}^{N} {\mathcal V}(|\vec{x}_{n}-\vec{x}_{m}|)$.
Then the    Hamilton function is specified by
\begin{equation}
 H(\{\vec{p}_{n} \},\{\vec{x}_{n}\})=\sum_{n=1}^{N}\left[\frac{\vec{p}_{n}^2}{2m}+\mathcal{U}(z_{n})\right] + V(\{\vec{x}_{n}\}).
\end{equation}
Throughout this paper the microscopic dynamics are assumed to be Newtonian generated by the Hamiltonian $H(\{\vec{p}_{n} \},\{\vec{x}_{n}\})$, in particular, collisions with the flat hard walls are elastic thereby conserving momentum parallel to the walls. The trajectory in the $N$ particle phase space is denoted by $(\{\vec{r}_{n}(t)\} ,\{ z_{n} (t)\},\{\vec{P}_{n}(t)\}, \{P_{n}^{z}(t)\})$ and all calculations are performed in the canonical ensemble. The thermodynamic limit $N\to \infty, A\to \infty$ is anticipated for fixed 2D number density $n_0=N/A$. With the accessible volume of the particles  $V=AL$, we find for the 3D number density $n=N/V=n_{0}/L$.

The modulation of the equilibrium density profile $n(z)$ in the slit is encoded in discrete Fourier components
\begin{equation}\label{eq:den}
 n_\mu = \int \diff z \exp( \text{i} Q_\mu z) n(z),
\end{equation}
where the mode index $\mu$ refers to  discrete wave numbers $Q_{\mu}=2\pi \mu /L$, $\mu \in \mathbb{Z}$ and integration is performed over the accessible slit $z\in [-L/2,L/2]$.

The fundamental variable of interest is the microscopic fluctuating density mode
\begin{equation}
\rho_{\mu}(\vec{q},t)=\sum_{n=1}^{N} \exp[\text{i} Q_\mu z_n (t)] \, \text{e}^{\text{i} \vec{q} \cdot \vec{r}_n(t)},
\end{equation}
where $\vec{q}=(q_{x},q_{y})$ are the conventional discrete (for finite $A$) wave vectors in the $x$-$y$-plane.
Particle number conservation relates the time derivative of $\rho_{\mu}(\vec{q},t)$ to the current densities parallel and perpendicular to the walls~\cite{Lang:2012}:
\begin{equation}
j_{\mu}^{\alpha}(\vec{q},t)\!=\! \frac{1}{m} \!\sum\limits_{n=1}^{N}b^{\alpha}(\hat{\vec{q}} \cdot {\vec{P}}_{n}(t),P_{n}^{z}(t)) \exp[\text{i} Q_\mu z_n (t)] \, \text{e}^{\text{i} \vec{q} \cdot \vec{r}_n(t)} ,
\end{equation}
with channel index $\alpha=\parallel,\perp$. Here, the short-hand notation for the unit vector $\hat{\vec{q}}=\vec{q}/q$ and  the selector $b^{\alpha}(x,z)=x \delta_{\alpha\parallel}+z\delta_{\alpha\perp}$ has been employed.
The emergence of several decay channels is reminiscent of the mode-coupling theory of molecules~\cite{Scheidsteger:1997,Franosch:1997}, where generalized density modes couple to both translational and rotational currents.

The basic quantity of the MCT of liquids in slit geometry is the generalization of the intermediate scattering function
\begin{equation}
S_{\mu\nu}(q,t)=\frac{1}{N} \langle\rho_{\mu}(\vec{q},t)^* \rho_{\nu}(\vec{q},0)  \rangle.
\end{equation}
We shall make use of a natural matrix notation  $[\bm{S}(q,t)]_{\mu\nu}=  S_{\mu\nu}(q,t)$, and similarly for other correlation functions throughout this paper.
MCT requires the Fourier coefficients $n_{\mu}$, the static structure factors $S_{\mu\nu}(q)=S_{\mu\nu}(q,t=0)$ and the static current-current correlators
\begin{align}\label{eq:J}
 [\bm{\mathcal{J}}(q)]^{\alpha\beta}_{\mu\nu}&=\mathcal{J}_{\mu\nu}^{\alpha \beta}(q)\nonumber\\
&=\frac{1}{N} \langle  j_{\mu}^{\alpha }(\vec{q})^* j_{\nu}^{\beta}(\vec{q})\rangle\nonumber\\
&=\delta^{\alpha\beta}v_\text{th}^2 \frac {n_{\mu-\nu}^*}{n_{0}},
\end{align}
as known input~\cite{Lang:2012} with thermal velocity $v_\text{th} = (k_B T /m)^{1/2}$.
General properties of the static and dynamic correlators including symmetry relations have been discussed in detail~\cite{Lang:2012}.

Employing the Zwanzig-Mori projection operator formalism~\cite{Forster:Hydrodynamic_Fluctuations,Goetze:Complex_Dynamics} exact equations of motion for the  collective correlators $S_{\mu \nu}(q,t)$ can be derived~\cite{Lang:2010,Lang:2012}. For later purposes it is more convenient to consider the Laplace-transformed equations, where the convention for the transformed  matrix correlators is $\hat{S}_{\mu\nu}(q,z)=\text{i}\int_{t=0}^{\infty}\diff t \, \text{e}^{\text{i} zt}S_{\mu \nu}(q,t)$, $\text{Im}[z]>0$~\footnote{The frequency $z$ should not be confused with the transversal variable $z$. The distinction is clear from the context.}.
The equations in the Laplace domain for $\hat{S}_{\mu\nu}(q,z)$ deal with generalized matrix-valued fraction representations. First, we express $\hat{S}_{\mu\nu}(q,z)$ in terms of current-memory kernels $\hat{K}_{\mu\nu}(q,z)$~\cite{Lang:2012},
\begin{equation}\label{eq:Ss}
\hat{\mathbf{S}}(q,z) = -\left[ z \mathbf{S}^{-1}(q) + \mathbf{S}^{-1}(q) \hat{\mathbf{K}}(q,z)\mathbf{S}^{-1}(q) \right]^{-1},
\end{equation}
which split by the perpendicular and parallel current to
\begin{equation}\label{eq:Ks}
 \hat{K}_{\mu\nu}(q,z) = \sum_{\alpha\beta=\parallel,\perp}b^{\alpha}(q,Q_{\mu})\hat{\mathcal{K}}^{\alpha\beta}_{\mu\nu}(q,z)
 b^{\beta}(q,Q_{\nu}).
\end{equation}
The kernel $[\hat{\bm{\mathcal K}}(q,z)]^{\alpha\beta}_{\mu\nu}=\hat{\mathcal{K}}^{\alpha\beta}_{\mu\nu}(q,z)$ can be represented by
\begin{equation}\label{eq:Kabs}
\hat{\bm{\mathcal K}}(q,z) = -\left[ z \bm{\mathcal J}^{-1}(q) + \bm{\mathcal J}^{-1}(q) \hat{\bm{\mathfrak M}}(q,z)\bm{\mathcal J}^{-1}(q) \right]^{-1},
\end{equation}
which involves the force kernel $\hat{\mathfrak{M}}_{\mu\nu}^{\alpha\beta}(q,z)$. The constitutive MCT ansatz expresses the force kernel in the time domain in terms of the intermediate scattering functions~\cite{Lang:2012},
\begin{align}\label{eq:Memory}
\mathfrak{M}^{\alpha\beta}_{\mu\nu}(q,t)& =  \frac{1}{2N^{3}} \sum_{\vec{q}_{1},\vec{q}_{2}}\sum_{\substack{\mu_{1}\mu_{2}\\ \nu_{1} \nu_{2}}}\mathcal{X}^{\alpha}_{\mu,\mu_{1}\mu_{2}}(\vec{q},\vec{q}_{1}\vec{q}_{2})\nonumber\\
&\times S_{\mu_{1}\nu_{1}}(q_{1},t) S_{\mu_{2}\nu_{2}}(q_{2},t) \mathcal{X}^{\beta}_{\nu,\nu_{1}\nu_{2}}(\vec{q},\vec{q}_{1}\vec{q}_{2})^*.
\end{align}
The vertices have been calculated relying on a systematic convolution approximation~\cite{Lang:2012},
\begin{align}\label{eq:Ypsilon}
\mathcal{X}^{\alpha}_{\mu,\mu_{1}\mu_{2}}&(\vec{q},\vec{q}_{1},\vec{q}_{2})\nonumber\\
 =& -N  v_\text{th}^2\frac{n_{0}}{L^2} \delta_{\vec{q},\vec{q}_{1}+\vec{q}_{2}}[b^{\alpha}(\hat{\vec{q}}\cdot\vec{q}_{1},Q_{\mu-\mu_{2}})\nonumber\\
&\times c_{\mu-\mu_{2},\mu_{1}}(q_{1})+(1\leftrightarrow 2)],
\end{align}
and involve the direct correlation functions $c_{\mu\nu}(q)$, which are related to the static structure factor $S_{\mu \nu} (q)$ by the inhomogeneous Ornstein-Zernike equation~\cite{Henderson:Fundamentals_of_inhomogeneous_fluids,Lang:2014b}.

The Eqs.~\eqref{eq:Ss}-\eqref{eq:Ypsilon} involve the initial conditions $S_{\mu\nu}(q,t=0)= S_{\mu\nu}(q)$ and $\mathcal{K}^{\alpha\beta}_{\mu\nu}(q,t=0) =  \mathcal{J}^{\alpha\beta}_{\mu\nu}(q)$ thereby constituting a complete set of coupled non-linear equations which have to be solved self-consistently. The investigations performed in the following rely on the proofs that the mode-coupling equations provide the existence of unique solutions, which has been demonstrated rigorously for Brownian dynamics in monocomponent simple liquids~\cite{Goetze:1995,Goetze:Complex_Dynamics} and mixtures~\cite{Franosch:2002} and only recently for the MCT with multiple relaxation channels, where the here discussed confined MCT is merely a special case~\cite{Lang:2013}.

\section{CONVERGENCE TO THE PLANAR MCT}\label{Sec:2DLimit}

In this section we discuss how the MCT equations converge towards the MCT for 2D liquids, which is one of the major results of this work. Let us emphasize that this requires  to discuss the theory for small but finite slit width $L>0$. Convergence implies that fluids confined to small slit widths behave similarly to  the truly two-dimensional case, the errors can be made arbitrarily small upon decreasing the slit width.

In the limiting regime of small  wall separation the in-plane motion is presumed to be close to the dynamics of a two-dimensional system. Furthermore, of all the structural properties entering the MCT equations  the two-dimensional structure factor should dominate the equations for small slit widths.
The MCT equations encode the confinement by the walls  via the structural input in terms of the density modes and the symmetry-adapted static structure factors both as initial values for the current correlators and for the intermediate scattering functions and on the associated direct correlation functions via the vertices in the MCT functional. One subtlety arises since for small wall separation $L>0$ the wave numbers $Q_\mu = 2\pi \mu/L$ associated with the perpendicular motion blow up.

\subsection{Density profile and static correlators}\label{Sec:2Dstatics}
Since the MCT relies on the structural properties of the fluid as known input, the question of convergence to a planar limit is intimately related to the structure of fluids in extreme confinement and the rapidity of the approach to a two-dimensional system. This issue has been addressed in two of our earlier works~\cite{Franosch:2012,Lang:2014a} in terms of a cluster and cumulant expansion in case of hard core and smooth potentials, respectively. There, one can not only estimate the order of convergence but actually calculate the leading corrections to both thermodynamic and structural quantities with respect to a planar reference system. Here we shall be interested only in the leading terms and quantify the order of  the corrections in terms of Landau symbols ${\cal O}(\cdot)$ and $o(\cdot)$ as the slit width becomes small $L\to 0$.

The convergence of the structural quantities has been demonstrated~\cite{Lang:2014a} assuming analytic wall potentials fulfilling the smoothness criterion ${\cal U}(z = \tilde{z}L)-{\cal U}(0) = {\cal O}(L)$,
which states that the particle-wall interaction should be controlled for fixed \emph{scaled} transverse coordinates $\tilde{z}$ as the slit width approaches zero. Then the density profile becomes flat~\cite{Lang:2014a} even on the scale of the plate distance $n(z) = (n_0/L) [1 + {\cal O}(L)]$.  This in turn implies convergence for the Fourier modes of the density,
\begin{align}\label{eq:nv}
n_{\mu}=
\begin{cases}
n_0=\text{const.} & \text{for } \mu =0, \\
{\cal O}(L) & \text{else}.
\end{cases}
\end{align}
Note, that the Fourier mode $n_{0}$ is independent of $L$ from the sum rule $n_{0}=\int\diff z n(z)\equiv N/A$.
We assume here the worst case of asymmetric walls, whereas for symmetric walls $\mathcal{O}(L)$ can be replaced by the faster convergence $\mathcal{O}(L^2)$, see Ref.~\cite{Lang:2014a}.
Let us repeat a word of warning here. The existence of the convergence is not guaranteed in general and, in particular, potentials diverging at the walls, e.g., Lennard-Jones and Coulomb potentials, do not belong to the class indicated above. Yet, pure hard walls trivially fulfill the smoothness criterion and they constitute the reference case we wish to address.
From the convergence properties of the density modes, Eq.~\eqref{eq:nv}, one can immediately infer the convergence of the static current correlator, Eq.~\eqref{eq:J}, to
\begin{equation}\label{eq:Jlim}
\mathcal{J}^{\alpha\beta}_{\mu \nu}(q)=\delta^{\alpha\beta}
\begin{cases}
  v_\text{th}^2 & \text{for } \mu=\nu,  \\
 {\cal O}(L) & \text{else},
\end{cases}
\end{equation}
which becomes to leading order diagonal with respect to the discrete modes.

Similarly, for the static structure factors it has been shown~\cite{Lang:2014a} that
\begin{equation}\label{eq:Slim}
S_{\mu \nu}(q)=
\begin{cases}
 S(q) [1+ {\cal O}(L^2)] & \text{for } \mu=\nu=0,  \\
(1-\delta_{\mu 0} ) \delta_{\mu\nu} +{\cal O}(L) & \text{else},
\end{cases}
\end{equation}
where $S(q)$ denotes the static structure factor of the corresponding 2D liquid. Thus the structure factor matrix $[\mathbf{S} (q)]_{\mu\nu}=S_{\mu \nu}(q)$ becomes  diagonal as well in the 2D limit and in addition, $S_{\mu\mu}(q)$  for $\mu\neq0$  becomes ideal-gas-like for $L\to 0$.

The  direct correlation function of the confined liquid in the slit converges to
\begin{equation}\label{eq:clim}
c_{\mu\nu}(q) = L^2 [c(q) \delta_{\mu 0}\delta_{\nu 0}+ L^2 \tilde{c}_{\mu\nu}(q)+o(L^2)],
\end{equation}
where $c(q)$ is the corresponding 2D direct correlation function and the correction amplitude $\tilde{c}_{\mu\nu}(q)$ is independent of the slit width $L$. The prime observation of Ref.~\cite{Lang:2014a} was that the  corrections are ${\cal O}(L^2)$ irrespective of the wall potential, whereas  for the structure factors for $(\mu,\nu) \neq (0,0)$ or the density profile the leading corrections are ${\cal O}(L)$. This latter property plays an important role for the convergence of MCT for confined liquids.

\subsection{Mode-coupling theory: $t$ finite, $L \to 0$}\label{Sec:t_fin}
We start the investigation of the convergence of the confined MCT towards the planar MCT by discussing the memory kernels. We assume first, that times $t$ and frequencies $z$ are held fixed independent of the wall separation, while the limit $L\to 0$ is performed.  For this purpose, it is favorable to introduce rescaled modes $\tilde{Q}_{\mu}=LQ_{\mu}=\mathcal{O}(L^0)$, which are independent of the slit size. Using the convergence of the direct correlation function, Eq.~\eqref{eq:clim}, neglecting terms of $\mathcal{O}(L^2)$ from Eq.~\eqref{eq:Ypsilon}, one obtains
\begin{align}\label{eq:Ypsilonex}
\mathcal{X}^\alpha_{\mu,\mu_1\mu_2}&(\vec{q},\vec{q}_1,\vec{q}_2)\nonumber\\
=&-N v_\text{th}^2 n_0 \delta_{\vec{q},\vec{q}_1+\vec{q}_2}\Big\{ \delta^{\alpha \parallel}(\hat{\vec{q}} \cdot\vec{q}_1) c(q_1) \delta_{\mu_1 0} \delta_{\mu_2 \mu}\nonumber\\
&+L \delta^{\alpha \perp} \tilde{Q}_{\mu-\mu_{2}} \tilde{c}_{\mu-\mu_{2},\mu_{1}}(q_{1})+(1\leftrightarrow 2)\Big\}.
\end{align}
Thereby, we find that  the leading contribution of the vertices in the limit $L\to 0$ stems from $\alpha=\parallel$. Keeping now only the leading order, i.e., we set $L=0$, and assuming a priori that $S_{\mu\nu}(q,t)=\mathcal{O}(L^0)$ for all $(\mu,\nu)$, in the 2D thermodynamic limit $N\to \infty$, $A\to \infty$ with $n_{0}=N/A$ fixed,  the nonvanishing memory kernel assumes the following form
\begin{align}\label{eq:MCT_functionals}
\mathfrak{M}^{\parallel\parallel}_{\mu\nu}(q,t) =& \frac{n_0 }{2}v_\text{th}^4 \int \frac{\diff^2 q_1}{(2\pi)^2} \big\{ [\hat{\vec{q}}\cdot \vec{q}_1 c(q_1)]^2 S_{00}(q_1,t) S_{\mu\nu}(q_2,t)
\nonumber \\ & + [\hat{\vec{q}}\cdot \vec{q}_1 c(q_1)] [\hat{\vec{q}}\cdot \vec{q}_2 c(q_2)] S_{0\nu}(q_1,t) S_{\mu 0}(q_2,t)\nonumber\\
& +  (1\leftrightarrow 2)
 \big\},
\end{align}
where $\vec{q}_2 = \vec{q}-\vec{q}_1$. In particular, for $\mu=\nu=0$ the memory kernel contains only couplings of $S_{00}(q,t)$
\begin{align}
& \mathfrak{M}^{\parallel\parallel}_{00}(q,t) =  \int \frac{\diff^2 q_1}{(2\pi)^2} V(\vec{q},\vec{q}_1\vec{q}_2) S_{00}(q_1,t) S_{00}(q_2,t),
\end{align}
and the  vertices $V(\vec{q},\vec{q}_1\vec{q}_2)$  coincide with the ones of the two-dimensional theory~\cite{Bayer:2007}
\begin{align}
&V(\vec{q},\vec{q}_{1},\vec{q}_{2})=\frac{n_{0}}{2 }
v_\text{th}^4
[(\hat{\vec{q}}\cdot \vec{q}_{1})c(q_{1})+(1\leftrightarrow 2) ]^2.
\end{align}
The  remaining memory kernels assuming $S_{\mu\nu}(q,t)=\mathcal{O}(L^0)$ for \emph{all} $(\mu,\nu)$ are of higher order in $L$, viz.
\begin{align}\label{eq:upperbound}
\mathfrak{M}^{\parallel \perp}_{\mu\nu}(q,t)& = \mathcal{O}(L),\nonumber\\
\mathfrak{M}^{\perp \parallel}_{\mu\nu}(q,t)&=\mathcal{O}(L),\nonumber\\
\mathfrak{M}^{\perp \perp}_{\mu\nu}(q,t) &=\mathcal{O}(L^{2}),
\end{align}
 as one infers from Eq.~\eqref{eq:Ypsilonex}.

The notable property of the MCT functional, Eqs.~\eqref{eq:MCT_functionals}, is that if the intermediate scattering functions $S_{\mu\nu}(q,t)$ are diagonal in the mode indices, this property is preserved by $\mathfrak{M}^{\parallel\parallel}_{\mu\nu}(q,t)$. Since also the static structure factors and current correlators, which serve as initial conditions, are diagonal to lowest order in $L$
the equations of motion, Eqs.~\eqref{eq:Ss}$-$\eqref{eq:Kabs} do not generate off-diagonal terms. More formally one can show that all time derivatives $\diff^l S_{\mu\nu}(q,t)/\diff t^l|_{t=0}, l\in \mathbb{N}_{0}$ are diagonal, similar to Ref.~\cite{Schilling:2002}. Since the solutions  have been demonstrated  to be unique~\cite{Lang:2013}, the thus constructed solution remains diagonal for all times $t>0$.

In particular, one finds that the equations of motion for $S_{00}(q,t)$ decouple completely from the remaining diagonal ones as we demonstrate below. To simplify notation we drop the mode indices and write $S(q,t) = S_{00}(q,t)$ and similarly for the static structure factor $S(q)=S_{00}(q,t=0)$. Furthermore we make contact with the notation of Ref.~\cite{Bayer:2007}:
\begin{equation}
\mathfrak{M}^{\parallel\parallel}_{00}(q,t) = \Omega_q^2 v_\text{th}^2 m(q,t),
\end{equation}
with the characteristic frequency $\Omega_{q}^2= q^2 v_\text{th}^2/ S(q)$.
Using this result and  Eq.~\eqref{eq:Jlim} it follows from the second fraction representation, Eq.~\eqref{eq:Kabs}, that its solutions are diagonal in $(\alpha,\beta)$  and in $(\mu,\nu)$,
and that the equations for $\hat{\mathcal{K}}^{\parallel\parallel}_{\mu\mu}(q,z)$  and $\hat{\mathcal{K}}^{\perp\perp}_{\mu\mu}(q,z)$ decouple for all $\mu$. Consequently, one obtains
 a closed equation for $\hat{\mathcal{K}}_{00}^{\parallel \parallel}(q,z)$  involving $\hat{m}(q,z)$, only. Abbreviating the 2D relaxation kernel  $\hat{K}(q,z)=\hat{K}_{00}(q,z) = q^2 \hat{\mathcal{K}}_{00}^{\parallel \parallel}(q,z)$ one finds from Eq.~\eqref{eq:Kabs}
\begin{equation}\label{eq:Ksfrac}
\hat{K}(q,z)=-\frac{q^2v^2_{\text{th}}}{z+\Omega_{q}^2\hat{m}(q,z)}.
\end{equation}
Taking advantage of the diagonality of $S_{\mu\nu}(q)$ and Eq.~\eqref{eq:Ss} one obtains the well-known double-fraction representation
\begin{equation}\label{eq:Ssfrac}
\hat{S}(q,z)=\frac{-S(q)}{z-\Omega^2_{q}/\left[z+\Omega^2_{q}\hat{m}(q,z)\right]}.
\end{equation}
Going back to the temporal domain leads to the generalized harmonic oscillator equation
\begin{equation}\label{eq:2DMCT}
 \ddot{S}(q,t)+\Omega_q^2 S(q,t) + \Omega_q^2 \int_0^t m(q,t-t')\dot{S}(q,t')\diff t'=0,
\end{equation}
coinciding  with the MCT equation of a 2D liquid~\cite{Bayer:2007}.

It is interesting to ask what the MCT equations yield for the remaining diagonal correlators $S_{\mu\mu}(q,t)$ for $\mu\neq 0$ for small plate separation $L$. While the equations for $S_{00}(q,t)$ allow for a direct limit $L=0$, the $L$-dependence cannot totally be eliminated for the other quantities. We evaluate Eq.~\eqref{eq:Kabs} again for $L=0$, and the contraction with the selectors, Eq.~\eqref{eq:Ks}, implies for the current kernel
$\hat{K}_{\mu\mu}(q,z)=-\tilde{Q}^2_{\mu}L^{-2}v^2_{\text{th}}/z$. The transversal coherent scattering function then becomes
\begin{equation}
\hat{S}_{\mu\mu}(q,z)=\frac{-1}{z-\tilde{Q}^2_{\mu}L^{-2}v^2_{\text{th}}/z} \qquad \text{for}\,\,\mu\neq 0,
\end{equation}
which is the equation of motion of uncoupled undamped oscillations. Thus for $L\to 0$ and $\mu\neq 0$ the correlators $S_{\mu\mu}(q,t)$
 display  fast harmonic oscillations with frequency $\Omega_{\mu} =\tilde{Q}_{\mu}L^{-1} v_\text{th} :=4\pi\mu/\tau(L,T)$ and are independent of the planar wave number $q$. Their basic period
$\tau=\tau(T,L)= 2L/v_\text{th}$ is just the time a single particle  with thermal velocity $ v_\text{th}$ needs for a single bounce between both walls. This time scale becomes small for $L\to0$, thus it is separated from the microscopic dynamics $\tau\ll t_{0}=1/\Omega_{q}$ of the planar dynamics. Indeed, if one inspects Eq.~\eqref{eq:Ssfrac} on the time scale $\tau$, i.e., $z\tau =\mathcal{O}(1)$, the planar correlator $S(q,t)$ for $t/\tau=\mathcal{O}(1)$ has not evolved for $L\to0$: $S(q,t)=S(q)$. On this rapid scale the MCT kernels are negligible, which is in accordance with the notion that they are designed to describe the slow dynamical processes leading to structural arrest. If one performs the same reasoning as described above neglecting the memory kernels for the MCT equations of motion of the tagged-particle correlator in Ref.~\cite{Lang:2014b}, then one obtains the same equation of motion for the transversal dynamics $\hat{S}^{(s)}_{\mu\mu}(q,z)=-1/\left[z-\tilde{Q}^2_{\mu}L^{-2}v^2_{\text{th}}/z\right] \qquad \text{for}\,\,\mu\neq 0$, if the tagged particle is of the same species as the host-liquid particles. Thus, the fast transversal dynamics of the collective intermediate scattering function reduces to the dynamics of the incoherent scattering function, i.e., $\hat{S}^{(s)}_{\mu\mu}(q,z)\equiv \hat{S}_{\mu\mu}(q,z)$ for $\mu\neq 0$ and $L\to 0$ indicating that the residual perpendicular dynamics is merely a one-particle dynamics.

We conclude that the equations of motion are capable to account for  a dynamical decoupling of lateral and perpendicular degrees of freedom, splitting off the planar glassy dynamics from  a fast one-dimensional ideal-gas like motion in a finite box. Nevertheless, we expect that the true transversal dynamics is not correctly contained in the MCT equations, but the fundamental time scales are still properly reflected.

\subsection{Mode-coupling theory: $L$ finite, $t \to \infty$ }
The present case allows to study the glass form factors as a function of the wall separation $L$.
Beyond a certain critical point, the MCT equations describe structural arrest characterized by nonvanishing glass-form factors, i.e., for confined liquids $F_{\mu\nu}(q):=\lim_{t\to\infty}S_{\mu\nu}(q,t)\neq0$. In experiments and simulations, these frozen-in parts describe the plateau values of the intermediate scattering function in the dense or supercooled regime~\cite{Megen:1994, Kob:1994, *Kob:1995a, *Kob:1995b,Goetze:1999}.
We  inspect the solutions for the glass-form factors in the limit $L\to 0$. From Eq.~\eqref{eq:2DMCT} one can readily extract the limit $\lim_{t\to \infty}\lim_{L\to 0}S_{00}(q,t)\equiv F(q)$, which coincides with the glass-form factors of the planar MCT~\cite{Bayer:2007} in case of structural arrest.
Interchanging the limits, i.e., taking first the limit $t\to\infty$ and then $L\to 0$, $\lim_{L\to 0}\lim_{t\to\infty}S_{\mu\nu}(q,t)=\lim_{L\to 0}F_{\mu \nu}(q)$, these glass-form factors  differ qualitatively from those obtained  from MCT in 2D~\cite{Bayer:2007}, as will be demonstrated in the following.

The fixed-point equation for glass-form factors $F_{\mu\nu}(q)$ have been derived in Ref.~\cite{Lang:2012} by performing the limit $z\to 0$. Since frozen-in force kernels display a pole at zero frequency,  $\mathfrak{M}^{\alpha\beta}_{\mu\nu}(q,z)=-(\mathfrak{F}^{\alpha\beta}_{\mu\nu}(q)/z)[1+o(1)]$ for $z\to 0$ one obtains
\begin{align}\label{eq:F}
 \mathbf{F}(q) =& \left[ \mathbf{S}^{-1}(q) + \mathbf{S}^{-1}(q) \mathbf{N}^{-1}(q) \mathbf{S}^{-1}(q) \right]^{-1},
\end{align}
with contractions
\begin{align}\label{eq:G}
[\bm{N}^{-1}]_{\mu\nu}(q)=& \sum_{\alpha\beta =\parallel,\perp} b^\alpha(q,L^{-1}\tilde{Q}_\mu)\nonumber\\
&\times[\bm{\mathcal{N}}^{-1}(q)]_{\mu\nu}^{\alpha\beta}
b^\beta(q,L^{-1}\tilde{Q}_\nu),
\end{align}
and the inverse of the frozen-in part of the MCT functional
\begin{equation}\label{eq:Ninv}
 [\bm{\mathcal{N}}^{-1}(q)]^{\alpha\beta}_{\mu\nu}:= [\bm{\mathcal J}(q)\bm{\mathfrak{F}}^{-1}(q)\bm{\mathcal J}(q)]^{\alpha\beta}_{\mu\nu}.
\end{equation}
It has been proven that these equations exhibit a unique maximal solution~\cite{Lang:2012,Lang:2013}.

The goal here is to find a self-consistent solution for the glass-form factors $F_{\mu\nu}(q)$ in terms of estimates in powers of $L$. The strategy is to perform a convergent iteration suggested in Ref.~\cite{Lang:2012} and to keep track of the respective orders in $L$. We initialize the iteration with $F_{\mu\nu}(q)=\mathcal{O}(L^0)$ for all $\mu,\nu$. Then, with the vertices, Eq.~\eqref{eq:Ypsilonex}, one infers for the long-time limits $\lim_{t\to \infty} \mathfrak{M}^{\alpha\beta}_{\mu\nu}(q,t)=\mathfrak{F}^{\alpha\beta}_{\mu\nu}(q)$, that the mode-coupling functionals display orders $\mathfrak{F}^{\parallel\parallel}_{\mu\nu}(q)=\mathcal{O}(L^0)$, $\mathfrak{F}^{\perp\parallel}_{\mu\nu}(q)=\mathcal{O}(L)$, $\mathfrak{F}^{\parallel\perp}_{\mu\nu}(q)=\mathcal{O}(L)$ and $\mathfrak{F}^{\perp\perp}_{\mu\nu}(q)=\mathcal{O}(L^2)$. In contrast to the frequency-dependent equations of motion, Eq.~\eqref{eq:Kabs}, the limit $L=0$ cannot be performed, since then the force-kernel matrix $\bm{\mathfrak{F}}(q)$ becomes singular and inversion in Eq.~\eqref{eq:Ninv} is not possible.

Keeping the leading-order estimates for the kernels one obtains immediately by Eqs.~\eqref{eq:F}$-$\eqref{eq:Ninv} the estimates  $F_{00}(q)=\mathcal{O}(L^0)$, $F_{0\nu}(q)=\mathcal{O}(L^2)$ ($\nu\neq 0$), $F_{\mu0}(q)=\mathcal{O}(L^2)$ ($\mu\neq 0$) and $F_{\mu\nu}(q)=\mathcal{O}(L^4)$ ($\mu,\nu\neq 0$) as first iterate. Reinserting the first iterate in the mode-coupling functional does not reduce the orders further, as  demonstrated in detail in Appendix. Hence the solutions,
\begin{align}\label{eq:glass-form}
 F_{00}(q)&=\mathcal{O}(L^0),\\\nonumber
F_{\mu 0}(q)&=\mathcal{O}(L^2)\qquad \text{for} \qquad\mu\neq 0, \\\nonumber
F_{0\nu}(q)&=\mathcal{O}(L^2)\qquad \text{for} \qquad\nu\neq 0, \\\nonumber
F_{\mu\nu}(q)&=\mathcal{O}(L^4)\qquad \text{for}\qquad \mu,\nu\neq 0,
\end{align}
are the unique solutions characterizing the glass states in extreme confinement for small but finite $L$.
In fact the Landau symbols ${\cal O}(\cdot)$ can be  replaced by asymptotic proportionality $\sim (\cdot )$, i.e., the orders cannot be improved.

Let us emphasize, that for all estimates $\lim_{L\to 0}F_{\mu\nu}(q)$, the frozen-in parts of the force kernels $\mathfrak{F}^{\alpha\beta}_{\mu\nu}(q)$ for $\alpha,\beta=\parallel\perp$ are mutually coupled. Thus, the equations for  $t\to\infty$ first do not decouple for small wall separations, in striking contrast to the case where the limit $L\to 0$ is performed for fixed finite times. In particular, this observation entails for the glass-form factor $\lim_{L\to 0}F_{00}(q)$, that  it does not coincide with the glass-form factors of the two-dimensional MCT, Eq.~\eqref{eq:2DMCT}; see the discussion in Appendix below Eq.~\eqref{eq:inv}. 


In conclusion, we have demonstrated, that the equations of motion display a delicate dependence in the limits $L\to 0$ and $t \to \infty$. Both limits do not commute, i.e., $\lim_{t\to \infty}\lim_{L\to 0}S_{\mu\nu}(q,t)\neq \lim_{L\to 0}\lim_{t\to \infty}S_{\mu\nu}(q,t)$.
This result necessarily implies the existence of an $L$-dependent diverging time scale $\tau_{L}$, on which the lateral dynamics couples to the transversal one.
We emphasize that the iteration scheme for the nonergodicity parameter, see Ref.~\cite{Lang:2012,Lang:2013}, remains valid for arbitrarily small $L>0$ and yields always the solutions, where lateral and transversal degrees of freedom are coupled.

\section{CONVERGENCE TO THE THREE-DIMENSIONAL BULK}\label{Sec:3DLimit}
In this subsection, we demonstrate that the MCT for confined fluid approaches the standard MCT of the 3D glass transition for wall separations approach infinity provided the static correlations are short-ranged.
 We discuss first the convergence of the static structure factors and the density profile. Then, we use these properties to extract the bulk limit of the MCT for confined liquids.

\subsection{Density profile and correlators}
We assume that the fluid-fluid and  fluid-wall interactions are  short-ranged and show that the bulk behavior for the average density and the two-particle static correlation functions is attained as the wall separation becomes large. This limit has to be performed such that the 3D density $n = n_0/L$ remains fixed.
The spatial dependence of the density  $n(z)$ arises due to the wall potential ${\cal U}(z)$ and the pair-potential $\mathcal{V}(|\vec{x}-\vec{x}'|)$. It displays a significant variation only in the vicinity of the wall, and becomes constant otherwise. Therefore we can write
\begin{equation}\label{eq:dendelta}
 n(z) = n + \Delta n(z),
\end{equation}
where $\Delta n(z)$ decays on a scale of a wall correlation length  $\xi$ away from the wall,  which can be defined by
\begin{equation}
 \xi = \frac{1}{n} \int \diff z |\Delta n (z)|.
\end{equation}
We recall that $n$ is the bulk density.
In general  $\xi$ depends on $L$, but for large wall separation $L\to \infty$ it assumes a finite limit depending on temperature and density, only. We assume that the confined liquid is sufficiently far away from a critical point.
 Then we obtain with Eq.~\eqref{eq:den} and $n_0=n L$
\begin{align}\label{eq:nmu3}
 n_\mu =& n_0 \left[ \delta_{\mu 0} + {\cal O}(\xi/L) \right],
\end{align}
since the integral [cf. Eq.~\eqref{eq:den}] over $\Delta n(z)$ is of order $\xi$.

Similar considerations apply for the distinct part $G^{(d)}(\vec{r},z,z')$ of the density-density correlation function $G(\vec{r},z,z')$ (see Ref.~\cite{Lang:2014a} for conventions). We write the distinct part as asymptotic expansion with respect to the corresponding bulk correlator
\begin{equation}\label{eq:Gd3lim}
 G^{(d)}(\vec{r},z,z') = \frac{1}{L}\left[G_{3\text{D}}^{(d)}(\vec{x})  + \Delta G^{(d)} (\vec{r},z,z')\right],
\end{equation}
where $\vec{x} = (\vec{r},z)$ and $G_{3\text{D}}^{(d)}(\vec{x})$ denotes the bulk distinct part of the density-density correlation function~\cite{Hansen:Theory_of_Simple_Liquids}, which displays rotational and translational invariance. The corrections  $\Delta G^{(d)}(\vec{r},z,z')$ again decay on a wall correlation length which we assume to be of same order as $\xi$.

The Fourier decomposition~\footnote{The convention for correlators in slit geometry is $A_{\mu \nu}(q)=\int\!\! \diff ^2 r \diff z \diff z'  A(\vec{r},z,z')\exp(-i Q_\mu z) \exp(i Q_\nu z')\text{e}^{-i \vec{q} \cdot \vec{r}} $} of the distinct and the corresponding self part [$G^{(s)}(\vec{r},z,z')=n(z) \delta(\vec{r})\delta(z-z')/n_{0}$] follows by Eqs.~\eqref{eq:nmu3} and~\eqref{eq:Gd3lim}
\begin{align}\label{eq:Sd}
S_{\mu\nu}^{(d)}(q)&= S^{(d)}(k) \delta_{\mu\nu} + {\cal O}(\xi/L),\nonumber\\
S_{\mu\nu}^{(s)}(q)&= \delta_{\mu\nu} + {\cal O}(\xi/L).
\end{align}
Here we adopt the convention that $\vec{k}= (\vec{q},Q_\mu)$ abbreviates the 3D wave vector, whereas
$\vec{q}$ is reserved for two-dimensional vectors, in particular $k = |\vec{k}| = (\vec{q}^2 + Q_\mu^2)^{1/2}$. For the structure factor we arrive at
\begin{equation}\label{eq:Slim3}
 S_{\mu\nu}(q) = S(k) \left[ \delta_{\mu\nu} + {\cal O}(\xi/L) \right],
\end{equation}
where $S(k)$ is the structure factor of the bulk liquid. From the Ornstein-Zernike relation in confined geometry~\cite{Henderson:Fundamentals_of_inhomogeneous_fluids,Lang:2014b} one immediately obtains
\begin{equation}\label{eq:c3}
 c_{\mu\nu}(q) = L [  c(k) \delta_{\mu\nu} + {\cal O}(\xi/L) ],
\end{equation}
with the bulk direct correlation function $ nc(k) =  1- 1/S(k)$~\cite{Hansen:Theory_of_Simple_Liquids}.
Thus, the static structure factors $S_{\mu\nu}(q)$ and the direct correlation matrix $(c_{\mu\nu}(q))$ become diagonal for $L\to \infty$.

\subsection{Mode-coupling theory}
For the bulk limit of the MCT equations of motion we also require the convergence of the static current correlator $\mathcal{J}_{\mu\nu}^{\alpha\beta}(q)$.  From Eqs.~\eqref{eq:J} and~\eqref{eq:nmu3} we obtain
\begin{equation}
 \mathcal{J}_{\mu\nu}^{\alpha\beta}(q) = \delta^{\alpha\beta} v_\text{th}^2 \delta_{\mu\nu} + {\cal O}(\xi/L).
\label{eq:J3dlimit}
\end{equation}
Thus all static correlators become diagonal with respect to the discrete mode indices  $\mu$ and $\nu$ in the bulk limit $L \to \infty$.
Substituting $c_{\mu\nu}(q)$ from Eq.~\eqref{eq:c3} into the memory kernel [Eq.~\eqref{eq:Memory}] and using Eq.~\eqref{eq:J3dlimit} we find
\begin{align}\label{eq:mem3}
 {\mathfrak M}_{\mu\nu}^{\alpha\beta}(q,t) = & \frac{1}{2 N } n^2 v_\text{th}^4 \sum_{\vec{q}_1} \sum_{\mu_1\nu_1} \nonumber \\
&\times \left[ b^\alpha\left(\frac{\vec{q}\cdot \vec{q}_1}{q} , Q_{\mu_1}) c(\vec{q}_1,Q_{\mu_1}\right) + (1 \leftrightarrow 2) \right] \nonumber \\
&\times \left[ b^\beta\left(\frac{\vec{q}\cdot \vec{q}_1}{q} , Q_{\nu_1}) c(\vec{q}_1,Q_{\nu_1}\right) + (1 \leftrightarrow 2) \right]^* \nonumber \\
&\times S_{\mu_1\nu_1}(q_1,t) S_{\mu_2\nu_2}(q_2,t),
\end{align}
where $\vec{q}_2 = \vec{q}-\vec{q}_1, \mu_2 = \mu-\mu_1, \nu_2 = \nu -\nu_1$. The memory kernel appears to be nondiagonal both in $(\alpha,\beta)$ and $(\mu,\nu)$.
However, as in the two-dimensional limit we shall show that the subspace  in which $S_{\mu\nu}(q,t)$ is diagonal, i.e.,
\begin{equation}\label{eq:diag}
S_{\mu\nu}(q,t)= S(k,t) \delta_{\mu\nu}
\end{equation}
 remains invariant under the MCT equations.

The assumption, Eq.~\eqref{eq:diag}, is consistent with the diagonality of the static structure factor $S_{\mu\nu}(q)$ [Eq.~\eqref{eq:Slim3}], the  initial condition for $S_{\mu\nu}(q,t)$. One infers for the memory kernel
 the simplification
\begin{align}
 {\mathfrak M}_{\mu\nu}^{\alpha\beta}(q,t)  =& \frac{1}{2 N} n^2 v_\text{th}^4
\sum_{\vec{k}_1 } \nonumber \\
&\times\left[ b^\alpha \left(\frac{\vec{q}\cdot \vec{q}_1}{q},Q_{\mu_1}\right) c(k_1) + (1 \leftrightarrow 2) \right] \nonumber \\
&\times \left[ b^\beta \left(\frac{\vec{q}\cdot \vec{q}_1}{q},Q_{\mu_1}\right) c(k_1) + (1 \leftrightarrow 2) \right] \nonumber \\
& \times S(k_1,t) S(k_2,t) \delta_{\mu\nu}
\label{eq:MCTsimplified}
\end{align}
with $\vec{k}_{2}=\vec{k}-\vec{k}_{1}$ and the corresponding wave vectors read $\vec{k}_i = (\vec{q}_i,Q_{\mu_i})$.

The thermodynamic limit $L\to \infty, A\to \infty, N\to \infty$ such that $N/A L = n$ restores isotropy in addition to homogeneity. Therefore we can choose the direction of the 'external' wave vector $\vec{k} = (\vec{q},Q_\mu)$ in a suitable way. Without restricting generality we choose $\vec{k} = (\vec{q}= \vec{0},Q_\mu)$, which implies for the projections $\hat{\vec{q}}\cdot\vec{q}_i = 0$. With $k=|\vec{k}| = |Q_\mu|$
 and $Q_{\mu_i}=\hat{\vec{k}}\cdot \vec{k}_i$,  the selector simplifies to
\begin{equation}
 b^{\alpha}(\hat{\vec{q}}\cdot\vec{q}_i, Q_{\mu_i}) = \hat{\vec{k}}\cdot\vec{k}_i \delta^{\alpha \perp}.
\label{eq:selector_simplified}
\end{equation}
Then the only nonvanishing elements of the memory kernel  in the thermodynamic limit become
\begin{equation}
 {\mathfrak M}_{\mu\nu}^{\perp\perp}(q,t) =    \delta_{\mu\nu} \int\frac{\diff^3 k_1}{(2\pi)^3} V(\vec{k}, \vec{k}_1 \vec{k}_2) S(k_1,t) S(k_2,t),
\label{eq:MCT3dlimit}
\end{equation}
with  the 3D vertices,
\begin{align}
V(\vec{k},\vec{k}_1\vec{k}_2) = \frac{n}{2 } v_\text{th}^4 \left[ (\hat{\vec{k}} \cdot \vec{k}_1) c(k_1) + (1 \leftrightarrow 2)\right]^{2}.
\end{align}
Again we make contact to established notation and write ${\mathfrak M}_{\mu\nu}^{\perp\perp}(q,t) =    \delta_{\mu\nu} \Omega_k^2 v_\text{th}^2 m(k,t)$
with $\Omega_k^2 = k^2 v_\text{th}^2 /S(k)$ and
$m(k,t)$ coincides with  the MCT kernel for a 3D liquid~\cite{Goetze:Complex_Dynamics}.

The remaining steps  are similar to the 2D case. Replacing  in Eq.~\eqref{eq:selector_simplified} $Q_{\mu_i}$ by $Q_{\mu}=k$, the total current correlators reduces to
 $\hat{K}_{\mu\nu}(q=0,z) = k^2 \hat{\mathcal{K}}^{\perp\perp}_{\mu\nu}(q=0,z)$ [see Eq.~\eqref{eq:Ks}].  By Eqs.~\eqref{eq:J3dlimit} and~\eqref{eq:Kabs} it becomes diagonal $\hat{K}_{\mu\nu}(q=0,z) = k^2 v_\text{th}^2 \delta_{\mu\nu}$ in the discrete wave numbers.
We abbreviate $\hat{K}(k,z) = \hat{K}_{\mu\mu}(q=0,z) = k^2 \hat{\mathcal{K}}^{\perp\perp}_{\mu\mu}(q=0,z)$ and arrive at the equation of motion
\begin{equation}\label{eq:Ksfrac3}
\hat{K}(q,z)=-\frac{k^2v^2_{\text{th}}}{z+\Omega_{k}^2\hat{m}(k,z)}.
\end{equation}
Diagonality of $S_{\mu\nu}(q)$ and of $\hat{K}_{\mu\nu}(q,z)$  in Eq.~\eqref{eq:Ss} implies that $\hat{S}_{\mu\nu}(q=0,z)=\hat{S}(k,z)\delta_{\mu\nu}$ remains diagonal, i.e., the assumption of Eq.~\eqref{eq:diag} remains consistent.  One obtains again a double-fraction representation
\begin{equation}\label{eq:Ssfrac3}
\hat{S}(k,z)=\frac{-S(k)}{z-\Omega^2_{k}/\left[z+\Omega^2_{k}\hat{m}(k,z)\right] }.
\end{equation}
Finally, the Laplace back transfrom of Eq.~\eqref{eq:Ssfrac3} yields the 3D generalized harmonic oscillator equation
\begin{equation}\label{eq:3DMCT}
 \ddot{S}(k,t)+\Omega_k^2 S(k,t) + \Omega_k^2 \int_0^t m(k,t-t')\dot{S}(k,t')\diff t'=0,
\end{equation}
which is the well-known MCT equation for glassy dynamics of a bulk liquid~\cite{Goetze:Complex_Dynamics}.
Since for large wall separations  we expect $S_{\mu\nu}(q,t)= {\cal O}(\xi/L) $ for $\mu\neq \nu$, we will not discuss the MCT equations for these correlators which describe the dynamics close to the walls.
In contrast to the 2D limit, the limits $t\to \infty$ and $L\to \infty$ do commute, since the fixed-point equation for the nonergodicity parameter, Eqs.~\eqref{eq:F}$-$\eqref{eq:Ninv}, converges properly for $L\to \infty$ to its bulk counterpart~\cite{Goetze:Complex_Dynamics}.

\section{SUMMARY AND CONCLUSIONS}\label{Sec:Summary}

We have demonstrated that the mode-coupling equations for confined liquids~\cite{Lang:2010,Lang:2012} for finite times converge to the MCT of the planar and bulk liquid for effective wall separation $L \to 0$ and $L \to \infty$, respectively.  In particular, we have shown how these limiting cases emerge from the reduction of the matrix-valued theory, which accounts for the inhomogeneous structure within the slit. In both cases, time is initially finite, while the limits $L\to 0$ and $L\to \infty$ are performed, respectively. The recovery of the correct limits demonstrates the consistency and robustness of the MCT ansatz also for  confined liquids.

Several conditions are imposed on the static level to guarantee the existence of these limits. For the 3D limit we have assumed short-ranged particle-particle  and particle-wall interactions and the thermodynamic state of the liquid is located sufficiently far away from a critical point. For the 2D limit, we  have required that the particle-wall interaction is such that structural properties approach their two-dimensional counterparts, in particular,
the density profile becomes flat sufficiently fast for $L\to 0$. The analyticity of the particle-wall interaction is sufficient for this condition.

The generalized matrix-valued intermediate scattering functions become diagonal in both limits. For finite times in the limit $L\to 0$ we have demonstrated that the 2D glassy dynamics of the lateral degrees of freedom decouples for $L\to 0$ from the transversal dynamics. While the dynamics parallel to the walls coincides with the 2D MCT dynamics, the latter display fast dynamics on a time scale $L/v_{\text{th}}$ mimicking ideal-gas-like motion in a one-dimensional box of size $L$. On this short time scale we find that the mode-coupling contributions to the transverse force fluctuations can be neglected for the description of this rapid dynamics. It becomes obvious that the transversal collective scattering function in this regime reduces to the incoherent scattering function indicating that this dynamics is a mere one-particle dynamics. More generally, we  expect that the MCT equations for the incoherent dynamics (see Ref.~\cite{Lang:2014b}) also converges to its planar and bulk case displaying similar features as found for the collective correlators discussed here.


The structural properties and thermodynamic phase behavior of extremely confined fluids can be determined using an effective two-dimensional pair potential, as has been shown recently~\cite{Franosch:2012,Lang:2014a}. Hence it would be interesting to use the two-dimensional MCT with these effective potentials and to compare to the corresponding results of the MCT in confinement. Similarly, one could compare computer simulations in slit geometry with two-dimensional simulations using the effective pair potential. An approach in the same spirit has been pursued for colloid-polymer mixtures where the effective Asakura-Oosawa depletion interaction has been employed also for studying dynamical properties. Similar strategies have been applied also for asymmetric colloidal mixtures~\cite{Zaccarelli:2004,Pham:2002} or star polymers~\cite{Foffi:2003}. We anticipate that this works reasonably well also in the present context although a microscopic justification is lacking.

%

Let us discuss the limit of small wall separations in more detail, since the planar limit displays peculiarities.   Interestingly, we find that the limits $t\to\infty$ and $L\to 0$ do not commute. If the limit $L\to0$ is performed first for fixed time, then applying the limit $t\to\infty$ leads to the glass-form factor of the corresponding 2D glass state. Taking the reverse order yields a glass state, where residual mutual couplings between the perpendicular and parallel frozen-in stresses remain. This subtle dependence on the performed limits suggests the existence of a divergent $L$-dependent time scale $\tau_L$, on which the transversal ideal-gas-like motion couples to the lateral degrees of freedom. This time scale occurs as a consequence of the non-commutativity of the limits $t\to \infty$ and $L\to0$ of the underlying MCT equations in confined geometry and provides an interesting prediction for future simulations and experiments.

Let us speculate on the relevance of such a  time scale $\tau_L$ diverging   for $L\to 0$ which
competes with the structural $\alpha$-relaxation time.
For instance, it becomes conceivable that for sufficiently small $L$ and fine-tuning of the density, the lateral degrees of freedom are frozen up to times $t \ll \tau_L$  and behave effectively as a  two-dimensional glass. At later times the coupling to the transversal degrees of freedom sets in and yields either a different glass state or even melts the glass entirely. Hence the transversal fluctuations effectively soften the planar interaction.
  Such a (partial) melting scenario
 differs from glass-glass transitions or reentrant transitions, e.g., for attractive glasses~\cite{Dawson:2000,Pham:2002,Eckert:2002} or
binary mixtures~\cite{Voigtmann:2011,Zaccarelli:2005,*Mayer:2008}, since it is a purely dynamical phenomenon for a single  thermodynamic state.
The emergence of two different regimes of the glassy dynamics is based on the assumption that the transversal degrees of freedom are slow variables even in extreme confinement. Therefore,
 it would be interesting to test this prediction experimentally or by computer simulation and verify that two different regimes, separated by a divergent time scale $\tau_L$, indeed behave qualitatively differently. Potential candidates for experimental realizations are superparamagnetic colloidal particles at a liquid surface~\cite{Deutschlaender:2014,Klix:2012,Mazoyer:2009,Koenig:2005} where small fluctuations in out-of-plane directions yield a generic mechanism to weakly couple to transversal degrees of freedom.

\begin{acknowledgments}
This work has been supported by the
Deutsche Forschungsgemeinschaft DFG via the  Research Unit FOR1394 ``Nonlinear Response to
Probe Vitrification''.
\end{acknowledgments}

\appendix
\begin{widetext}
\section{Glass-form factors in extreme confinement}
\label{app:form factors}
In this Appendix we investigate the convergence of the glass-form factors, i.e., $\lim_{L\to 0}F_{\mu\nu}(q)$.  We demonstrate here that the estimate   $F_{00}(q)=\mathcal{O}(L^0)$, $F_{0\nu}(q)=\mathcal{O}(L^2)$ ($\nu\neq 0$), $F_{\mu0}(q)=\mathcal{O}(L^2)$ ($\mu\neq 0$) and $F_{\mu\nu}(q)=\mathcal{O}(L^4)$ ($\mu,\nu\neq 0$) yields a consistent solution for the fixed-point equation; see Eqs.~\eqref{eq:F}$-$\eqref{eq:Ninv}.
The required ingredient for this analysis is the asymptotic behavior of the vertices for $L\to 0$, Eq.~\eqref{eq:Ypsilonex}. For all $\alpha ,\beta$, we list here the expanded functionals to lowest required powers in $L$ [convention $\vec{q}_{2}=\vec{q}-\vec{q}_{1}$]:

$\alpha=\beta=\parallel$:\\
\begin{align}
 \mathfrak{F}^{\parallel\parallel}_{\mu\nu}(q)=&\frac{v_{\text{th}}^4}{2N} \sum_{\vec{q}_{1}}\Big[(\hat{\vec{q}} \cdot\vec{q}_1)^2 c(q_1)^2 F_{00}(q_{1}) F_{\mu\nu}(q_{2})\nonumber\\
 &+(\hat{\vec{q}} \cdot\vec{q}_1)(\hat{\vec{q}} \cdot\vec{q}_2) c(q_1)c(q_{2}) F_{0\nu}(q_{1}) F_{\mu 0}(q_{2})+ (1\leftrightarrow 2)\Big]\nonumber\\
 &+L^2\frac{v_{\text{th}}^4}{2N} \sum_{\vec{q}_{1}}\sum_{\nu_1 \nu_{2}}\Big[ (\hat{\vec{q}}\cdot\vec{q}_1)^2 c(q_1)\tilde{c}_{\nu-\nu_{2},\nu_{1}}^*(q_{1})F_{0\nu_{1}}(q_{1})F_{\mu \nu_{2}}(q_{2})\nonumber\\
 &+(\hat{\vec{q}} \cdot\vec{q}_1)(\hat{\vec{q}} \cdot\vec{q}_2)c(q_{1})\tilde{c}_{\nu-\nu_{1},\nu_{2}}^*(q_{2}) F_{0\nu_{1}}(q_{1})F_{\mu \nu_{2}}(q_{2})+(1\leftrightarrow 2)\Big]\nonumber\\
 &+L^2\frac{v_{\text{th}}^4}{2N} \sum_{\vec{q}_{1}}\sum_{\mu_1 \mu_{2}}\Big[ (\hat{\vec{q}} \cdot\vec{q}_1)^2 c(q_1)\tilde{c}_{\mu-\mu_{2},\mu_{1}}(q_{1})F_{\mu_{1}0}(q_{1})F_{\mu_{2} \nu}(q_{2})\nonumber\\
 &+(\hat{\vec{q}} \cdot\vec{q}_1)(\hat{\vec{q}} \cdot\vec{q}_2)c(q_{1})\tilde{c}_{\mu-\mu_{1},\mu_{2}}(q_{2}) F_{\mu_{1}0}(q_{1})F_{\mu_{2} \nu}(q_{2})+(1\leftrightarrow 2)\Big],
 \end{align}
$\alpha=\parallel,\beta=\perp$:\\
\begin{align}
\mathfrak{F}^{\parallel\perp}_{\mu\nu}(q)
 =&L\frac{v_{\text{th}}^4}{2N} \sum_{\vec{q}_{1}}\sum_{ \nu_{1} \nu_{2}}(\hat{\vec{q}}\cdot\vec{q}_1) c(q_1) \tilde{Q}_{\nu-\nu_{2}} \tilde{c}_{\nu-\nu_{2},\nu_{1}}^*(q_{1})F_{0\nu_1}(q_{1}) F_{\mu \nu_{2}}(q_{2})\nonumber\\
 &+L\frac{v_{\text{th}}^4}{2N} \sum_{\vec{q}_{1}}\sum_{ \nu_{1} \nu_{2}}(\hat{\vec{q}}\cdot\vec{q}_1) c(q_1) \tilde{Q}_{\nu-\nu_{1}} \tilde{c}_{\nu-\nu_{1},\nu_{2}}^*(q_{2})F_{0\nu_1}(q_{1}) F_{\mu \nu_{2}}(q_{2})\nonumber\\
 &+(1\leftrightarrow 2),
\end{align}
$\alpha=\perp,\beta=\parallel$:\\
\begin{align}
\mathfrak{F}^{\perp\parallel}_{\mu\nu}(q)=&L\frac{v_{\text{th}}^4}{2N} \sum_{\vec{q}_{1}}\sum_{ \mu_{1} \mu_{2}}(\hat{\vec{q}}\cdot\vec{q}_1) c(q_1) \tilde{Q}_{\mu-\mu_{2}} \tilde{c}_{\mu-\mu_{2},\mu_{1}}(q_{1})F_{\mu_1 0}(q_{1}) F_{\mu_{2} \nu}(q_{2})\nonumber\\
 &+L\frac{v_{\text{th}}^4}{2N} \sum_{\vec{q}_{1}}\sum_{ \mu_{1} \mu_{2}}(\hat{\vec{q}}\cdot\vec{q}_2) c(q_2) \tilde{Q}_{\mu-\mu_{2}} \tilde{c}_{\mu-\mu_{2},\mu_{1}}(q_{1})F_{\mu_1\nu}(q_{1}) F_{\mu_{2}0}(q_{2})\nonumber\\
 &+(1\leftrightarrow 2).
\end{align}
$\alpha=\beta=\perp$:\\
\begin{align}
 \mathfrak{F}^{\perp\perp}_{\mu\nu}(q)=&L^2\frac{v_{\text{th}}^4}{2N} \sum_{\vec{q}_{1}}\sum_{\substack{\mu_{1}\mu_{2}\\ \nu_{1} \nu_{2}}}\tilde{Q}_{\mu-\mu_{2}} \tilde{Q}_{\nu-\nu_{2}}\tilde{c}_{\mu-\mu_{2},\mu_{1}}(q_{1}) \tilde{c}_{\nu-\nu_{2},\nu_{1}}^*(q_{1})F_{\mu_{1}\nu_{1}}(q_{1}) F_{\mu_{2}\nu_{2}}(q_{2})\nonumber\\
 &+L^2\frac{v_{\text{th}}^4}{2N} \sum_{\vec{q}_{1}}\sum_{\substack{\mu_{1}\mu_{2}\\ \nu_{1} \nu_{2}}}\tilde{Q}_{\mu-\mu_{2}} \tilde{Q}_{\nu-\nu_{1}}\tilde{c}_{\mu-\mu_{2},\mu_{1}}(q_{1}) \tilde{c}_{\nu-\nu_{1},\nu_{2}}^*(q_{2})F_{\mu_{1}\nu_{1}}(q_{1}) F_{\mu_{2}\nu_{2}}(q_{2})\nonumber\\
 &+ (1\leftrightarrow 2).
 \end{align}
We insert our estimates for the glass-form factors, Eq.~\eqref{eq:glass-form}, and count the leading orders in $L$. It turns out that for estimating the orders, it is sufficient to distinguish between
 mode index zero ('$0$') and a generic non-zero mode ('$\bar{0}$'). In this short-hand  matrix notation, we find
\begin{equation}
\left( \mathfrak{F}^{\alpha\beta}_{\mu\nu}(q)\right)=\bbordermatrix{
  & \parallel 0	& \parallel \bar{0} &\perp 0& \perp\bar{0}   \cr
\parallel 0& \mathcal{O}(L^0) & \mathcal{O}(L^2)& \mathcal{O}(L^3)&\mathcal{O}(L^1) \cr
\parallel \bar{0} & \mathcal{O}(L^2) & \mathcal{O}(L^4)& \mathcal{O}(L^5)  & \mathcal{O}(L^3)\cr
\perp 0 & \mathcal{O}(L^3) & \mathcal{O}(L^5)& \mathcal{O}(L^6)& \mathcal{O}(L^4)  \cr
\perp \bar{0} & \mathcal{O}(L^1) & \mathcal{O}(L^3) & \mathcal{O}(L^4) & \mathcal{O}(L^2)\cr
}.
\end{equation}
The inversion of this matrix can be done by introducing block matrices
\begin{equation}
\left( \mathfrak{F}^{\alpha\beta}_{\mu\nu}(q)\right)=\bbordermatrix{
  & \parallel	& \perp    \cr
\parallel & \bm{A}&  \bm{B}  \cr
\perp &  \bm{C} & \bm{D} \cr
},
\end{equation}
e.g.,
\begin{equation}
\bm{A}\equiv\left( A_{\mu\nu}\right)=\bbordermatrix{
  & 0	& \bar{0}    \cr
0 & \mathcal{O}(L^{0})&  \mathcal{O}(L^{2})  \cr
\bar{0} &  \mathcal{O}(L^{2}) & \mathcal{O}(L^{4}) \cr
},
\end{equation}
etc.
Block-matrix inversion yields
\begin{equation}\label{eq:inv}
\left( [\bm{\mathfrak{F}}^{-1}(q)]^{\alpha\beta}_{\mu\nu}\right)=\bbordermatrix{
  & \parallel	& \perp    \cr
\parallel & (\bm{A}-\bm{B}\bm{D}^{-1}\bm{C})^{-1}&  -\bm{A}^{-1}\bm{B}(\bm{D}-\bm{C}\bm{A}^{-1}\bm{B})^{-1}  \cr
\perp &  -\bm{D}^{-1}\bm{C}(\bm{A}-\bm{B}\bm{D}^{-1}\bm{C})^{-1} & (\bm{D}-\bm{C}\bm{A}^{-1}\bm{B})^{-1} \cr
}.
\end{equation}
An important insight is, that the estimates of each entry, e.g., compare $\bm{A}$ and $\bm{B}\bm{D}^{-1}\bm{C}$, are of the same order in $L$. The consequence is, that each entry of $\left( [\bm{\mathfrak{F}}^{-1}(q)]^{\alpha\beta}_{\mu\nu}\right)$ contains contributions from all force kernels $\mathfrak{F}^{\alpha\beta}_{\mu\nu}(q)$ with the various combinations of $\alpha$ and $\beta$. In particular, we emphasize that for the  entry $(\alpha=\beta=\parallel;\mu=\nu=0)$ the scalar
\begin{equation}\label{eq:coup1}
(\bm{B}\bm{D}^{-1}\bm{C})_{00}=\sum_{\kappa \gamma}\mathfrak{F}^{\parallel \perp}_{0\kappa}[(\bm{\mathfrak{F}}^{\perp\perp})^{-1}]_{\kappa\gamma} \mathfrak{F}^{\perp\parallel}_{\gamma 0}=\mathcal{O}(L^0),
\end{equation}
is of the same order as $A_{00}=\mathfrak{F}^{\parallel \parallel}_{0 0}$, the latter corresponding to the arrested 2D MCT kernel. This in turn implies that $A_{00}-(\bm{B}\bm{D}^{-1}\bm{C})_{00}$ does not coincide with the expected 2D result:
\begin{equation}\label{eq:coup2}
A_{00}-(\bm{B}\bm{D}^{-1}\bm{C})_{00}\neq \mathfrak{F}^{\parallel \parallel}_{0 0}.
\end{equation}

Explicit inversion leads to
\begin{align}
&\left( [\bm{\mathfrak{F}}^{-1}(q)]^{\alpha\beta}_{\mu\nu}\right)\\ \nonumber
&=\bbordermatrix{
  & \parallel 0	& \parallel \bar{0} &\perp 0& \perp\bar{0}   \cr
\parallel 0& \mathcal{O}(L^0) & \mathcal{O}(L^{-2})& \mathcal{O}(L^{-3})&\mathcal{O}(L^{-1}) \cr
\parallel \bar{0} & \mathcal{O}(L^{-2}) & \mathcal{O}(L^{-4})& \mathcal{O}(L^{-5})  & \mathcal{O}(L^{-3})\cr
\perp 0 & \mathcal{O}(L^{-3}) & \mathcal{O}(L^{-5})& \mathcal{O}(L^{-6})& \mathcal{O}(L^{-4})  \cr
\perp \bar{0} & \mathcal{O}(L^{-1}) & \mathcal{O}(L^{-3}) & \mathcal{O}(L^{-4}) & \mathcal{O}(L^{-2})\cr
}.
\end{align}
From the diagonality property of the current correlator, Eq.~\eqref{eq:Jlim}, $\left( [\bm{\mathcal{N}}^{-1}(q)]^{\alpha\beta}_{\mu\nu}\right)$ retains the same estimates [see Eq.~\eqref{eq:Ninv}].
Contraction with the selectors, Eq.~\eqref{eq:G}, yields
in matrix notation,
\begin{equation}
\left( [\bm{N}^{-1}]_{\mu\nu}(q)\right)=\bbordermatrix{
  & 0	& \bar{0}    \cr
0 & \mathcal{O}(L^{0})&  \mathcal{O}(L^{-2})  \cr
\bar{0} &  \mathcal{O}(L^{-2}) & \mathcal{O}(L^{-4}) \cr
}.
\end{equation}
Since the static structure factor is diagonal to leading order, see Eq.~\eqref{eq:Slim}, one infers from Eq.~\eqref{eq:F} the final result:
\begin{equation}
\left( F_{\mu\nu}(q)\right)=\bbordermatrix{
  & 0	& \bar{0}    \cr
0 & \mathcal{O}(L^{0})&  \mathcal{O}(L^{2})  \cr
\bar{0} &  \mathcal{O}(L^{2}) & \mathcal{O}(L^{4}) \cr
},
\end{equation}
where we have recovered our initial ansatz.
Since the solutions are unique~\cite{Lang:2013}, we have found the consistent solution for $L\to 0$. In particular, $F_{00}(q)$ couples to all residual kernels of $\mathfrak{F}^{\alpha\beta}_{\mu\nu}(q)$ [see explicitly Eqs.~\eqref{eq:coup1} and~\eqref{eq:coup2}] and thereby the such obtained glass-form factor differs from a purely two-dimensional glass state~\cite{Bayer:2007}.
\end{widetext}


\begin{thebibliography}{73}%
\makeatletter
\providecommand \@ifxundefined [1]{%
 \@ifx{#1\undefined}
}%
\providecommand \@ifnum [1]{%
 \ifnum #1\expandafter \@firstoftwo
 \else \expandafter \@secondoftwo
 \fi
}%
\providecommand \@ifx [1]{%
 \ifx #1\expandafter \@firstoftwo
 \else \expandafter \@secondoftwo
 \fi
}%
\providecommand \natexlab [1]{#1}%
\providecommand \enquote  [1]{``#1''}%
\providecommand \bibnamefont  [1]{#1}%
\providecommand \bibfnamefont [1]{#1}%
\providecommand \citenamefont [1]{#1}%
\providecommand \href@noop [0]{\@secondoftwo}%
\providecommand \href [0]{\begingroup \@sanitize@url \@href}%
\providecommand \@href[1]{\@@startlink{#1}\@@href}%
\providecommand \@@href[1]{\endgroup#1\@@endlink}%
\providecommand \@sanitize@url [0]{\catcode `\\12\catcode `\$12\catcode
  `\&12\catcode `\#12\catcode `\^12\catcode `\_12\catcode `\%12\relax}%
\providecommand \@@startlink[1]{}%
\providecommand \@@endlink[0]{}%
\providecommand \url  [0]{\begingroup\@sanitize@url \@url }%
\providecommand \@url [1]{\endgroup\@href {#1}{\urlprefix }}%
\providecommand \urlprefix  [0]{URL }%
\providecommand \Eprint [0]{\href }%
\providecommand \doibase [0]{http://dx.doi.org/}%
\providecommand \selectlanguage [0]{\@gobble}%
\providecommand \bibinfo  [0]{\@secondoftwo}%
\providecommand \bibfield  [0]{\@secondoftwo}%
\providecommand \translation [1]{[#1]}%
\providecommand \BibitemOpen [0]{}%
\providecommand \bibitemStop [0]{}%
\providecommand \bibitemNoStop [0]{.\EOS\space}%
\providecommand \EOS [0]{\spacefactor3000\relax}%
\providecommand \BibitemShut  [1]{\csname bibitem#1\endcsname}%
\let\auto@bib@innerbib\@empty
\bibitem [{\citenamefont {L\"owen}(2001)}]{Lowen:2001}%
  \BibitemOpen
  \bibfield  {author} {\bibinfo {author} {\bibfnamefont {H.}~\bibnamefont
  {L\"owen}},\ }\href {http://stacks.iop.org/0953-8984/13/i=24/a=201}
  {\bibfield  {journal} {\bibinfo  {journal} {J. Phys. Condens. Matter}\
  }\textbf {\bibinfo {volume} {13}},\ \bibinfo {pages} {R415} (\bibinfo {year}
  {2001})}\BibitemShut {NoStop}%
\bibitem [{\citenamefont {Schmidt}\ and\ \citenamefont
  {L\"owen}(1996)}]{Schmidt:1996}%
  \BibitemOpen
  \bibfield  {author} {\bibinfo {author} {\bibfnamefont {M.}~\bibnamefont
  {Schmidt}}\ and\ \bibinfo {author} {\bibfnamefont {H.}~\bibnamefont
  {L\"owen}},\ }\href {\doibase 10.1103/PhysRevLett.76.4552} {\bibfield
  {journal} {\bibinfo  {journal} {Phys. Rev. Lett.}\ }\textbf {\bibinfo
  {volume} {76}},\ \bibinfo {pages} {4552} (\bibinfo {year}
  {1996})}\BibitemShut {NoStop}%
\bibitem [{\citenamefont {Schmidt}\ and\ \citenamefont
  {L\"owen}(1997)}]{Schmidt:1997}%
  \BibitemOpen
  \bibfield  {author} {\bibinfo {author} {\bibfnamefont {M.}~\bibnamefont
  {Schmidt}}\ and\ \bibinfo {author} {\bibfnamefont {H.}~\bibnamefont
  {L\"owen}},\ }\href {\doibase 10.1103/PhysRevE.55.7228} {\bibfield  {journal}
  {\bibinfo  {journal} {Phys. Rev. E}\ }\textbf {\bibinfo {volume} {55}},\
  \bibinfo {pages} {7228} (\bibinfo {year} {1997})}\BibitemShut {NoStop}%
\bibitem [{\citenamefont {Alba-Simionesco}\ \emph {et~al.}(2006)\citenamefont
  {Alba-Simionesco}, \citenamefont {Coasne}, \citenamefont {Dosseh},
  \citenamefont {Dudziak}, \citenamefont {Gubbins}, \citenamefont
  {Radhakrishnan},\ and\ \citenamefont {Sliwinska-Bartkowiak}}]{Alba:2006}%
  \BibitemOpen
  \bibfield  {author} {\bibinfo {author} {\bibfnamefont {C.}~\bibnamefont
  {Alba-Simionesco}}, \bibinfo {author} {\bibfnamefont {B.}~\bibnamefont
  {Coasne}}, \bibinfo {author} {\bibfnamefont {G.}~\bibnamefont {Dosseh}},
  \bibinfo {author} {\bibfnamefont {G.}~\bibnamefont {Dudziak}}, \bibinfo
  {author} {\bibfnamefont {K.~E.}\ \bibnamefont {Gubbins}}, \bibinfo {author}
  {\bibfnamefont {R.}~\bibnamefont {Radhakrishnan}}, \ and\ \bibinfo {author}
  {\bibfnamefont {M.}~\bibnamefont {Sliwinska-Bartkowiak}},\ }\href
  {http://stacks.iop.org/0953-8984/18/i=6/a=R01} {\bibfield  {journal}
  {\bibinfo  {journal} {Journal of Physics: Condensed Matter}\ }\textbf
  {\bibinfo {volume} {18}},\ \bibinfo {pages} {R15} (\bibinfo {year}
  {2006})}\BibitemShut {NoStop}%
\bibitem [{\citenamefont {Klafter}\ and\ \citenamefont
  {Drake}(1989)}]{Klafter:Restricted}%
  \BibitemOpen
  \bibfield  {author} {\bibinfo {author} {\bibfnamefont {J.}~\bibnamefont
  {Klafter}}\ and\ \bibinfo {author} {\bibfnamefont {J.~M.}\ \bibnamefont
  {Drake}},\ }\href@noop {} {\emph {\bibinfo {title} {Molecular Dynamics in
  Restricted Geometry}}}\ (\bibinfo  {publisher} {Wiley},\ \bibinfo {address}
  {New York},\ \bibinfo {year} {1989})\BibitemShut {NoStop}%
\bibitem [{\citenamefont {Mittal}\ \emph {et~al.}(2008)\citenamefont {Mittal},
  \citenamefont {Truskett}, \citenamefont {Errington},\ and\ \citenamefont
  {Hummer}}]{Mittal:2008}%
  \BibitemOpen
  \bibfield  {author} {\bibinfo {author} {\bibfnamefont {J.}~\bibnamefont
  {Mittal}}, \bibinfo {author} {\bibfnamefont {T.~M.}\ \bibnamefont
  {Truskett}}, \bibinfo {author} {\bibfnamefont {J.~R.}\ \bibnamefont
  {Errington}}, \ and\ \bibinfo {author} {\bibfnamefont {G.}~\bibnamefont
  {Hummer}},\ }\href {\doibase 10.1103/PhysRevLett.100.145901} {\bibfield
  {journal} {\bibinfo  {journal} {Phys. Rev. Lett.}\ }\textbf {\bibinfo
  {volume} {100}},\ \bibinfo {pages} {145901} (\bibinfo {year}
  {2008})}\BibitemShut {NoStop}%
\bibitem [{\citenamefont {Nugent}\ \emph {et~al.}(2007)\citenamefont {Nugent},
  \citenamefont {Edmond}, \citenamefont {Patel},\ and\ \citenamefont
  {Weeks}}]{Nugent:2007}%
  \BibitemOpen
  \bibfield  {author} {\bibinfo {author} {\bibfnamefont {C.~R.}\ \bibnamefont
  {Nugent}}, \bibinfo {author} {\bibfnamefont {K.~V.}\ \bibnamefont {Edmond}},
  \bibinfo {author} {\bibfnamefont {H.~N.}\ \bibnamefont {Patel}}, \ and\
  \bibinfo {author} {\bibfnamefont {E.~R.}\ \bibnamefont {Weeks}},\ }\href
  {\doibase 10.1103/PhysRevLett.99.025702} {\bibfield  {journal} {\bibinfo
  {journal} {Phys. Rev. Lett.}\ }\textbf {\bibinfo {volume} {99}},\ \bibinfo
  {pages} {025702} (\bibinfo {year} {2007})}\BibitemShut {NoStop}%
\bibitem [{\citenamefont {Fehr}\ and\ \citenamefont
  {L\"owen}(1995)}]{Fehr:1995}%
  \BibitemOpen
  \bibfield  {author} {\bibinfo {author} {\bibfnamefont {T.}~\bibnamefont
  {Fehr}}\ and\ \bibinfo {author} {\bibfnamefont {H.}~\bibnamefont {L\"owen}},\
  }\href {\doibase 10.1103/PhysRevE.52.4016} {\bibfield  {journal} {\bibinfo
  {journal} {Phys. Rev. E}\ }\textbf {\bibinfo {volume} {52}},\ \bibinfo
  {pages} {4016} (\bibinfo {year} {1995})}\BibitemShut {NoStop}%
\bibitem [{\citenamefont {Krishnan}\ and\ \citenamefont
  {Ayappa}(2012)}]{Krishnan:2012}%
  \BibitemOpen
  \bibfield  {author} {\bibinfo {author} {\bibfnamefont {S.~H.}\ \bibnamefont
  {Krishnan}}\ and\ \bibinfo {author} {\bibfnamefont {K.~G.}\ \bibnamefont
  {Ayappa}},\ }\href {\doibase 10.1103/PhysRevE.86.011504} {\bibfield
  {journal} {\bibinfo  {journal} {Phys. Rev. E}\ }\textbf {\bibinfo {volume}
  {86}},\ \bibinfo {pages} {011504} (\bibinfo {year} {2012})}\BibitemShut
  {NoStop}%
\bibitem [{\citenamefont {Gallo}\ \emph
  {et~al.}(2000{\natexlab{a}})\citenamefont {Gallo}, \citenamefont {Rovere},\
  and\ \citenamefont {Spohr}}]{Gallo:2000a}%
  \BibitemOpen
  \bibfield  {author} {\bibinfo {author} {\bibfnamefont {P.}~\bibnamefont
  {Gallo}}, \bibinfo {author} {\bibfnamefont {M.}~\bibnamefont {Rovere}}, \
  and\ \bibinfo {author} {\bibfnamefont {E.}~\bibnamefont {Spohr}},\ }\href
  {\doibase 10.1103/PhysRevLett.85.4317} {\bibfield  {journal} {\bibinfo
  {journal} {Phys. Rev. Lett.}\ }\textbf {\bibinfo {volume} {85}},\ \bibinfo
  {pages} {4317} (\bibinfo {year} {2000}{\natexlab{a}})}\BibitemShut {NoStop}%
\bibitem [{\citenamefont {Gallo}\ \emph
  {et~al.}(2000{\natexlab{b}})\citenamefont {Gallo}, \citenamefont {Rovere},\
  and\ \citenamefont {Spohr}}]{Gallo:2000b}%
  \BibitemOpen
  \bibfield  {author} {\bibinfo {author} {\bibfnamefont {P.}~\bibnamefont
  {Gallo}}, \bibinfo {author} {\bibfnamefont {M.}~\bibnamefont {Rovere}}, \
  and\ \bibinfo {author} {\bibfnamefont {E.}~\bibnamefont {Spohr}},\ }\href
  {\doibase 10.1063/1.1328073} {\bibfield  {journal} {\bibinfo  {journal} {J.
  Chem. Phys.}\ }\textbf {\bibinfo {volume} {113}},\ \bibinfo {pages} {11324}
  (\bibinfo {year} {2000}{\natexlab{b}})}\BibitemShut {NoStop}%
\bibitem [{\citenamefont {Gallo}\ \emph {et~al.}(2009)\citenamefont {Gallo},
  \citenamefont {Attili},\ and\ \citenamefont {Rovere}}]{Gallo:2009}%
  \BibitemOpen
  \bibfield  {author} {\bibinfo {author} {\bibfnamefont {P.}~\bibnamefont
  {Gallo}}, \bibinfo {author} {\bibfnamefont {A.}~\bibnamefont {Attili}}, \
  and\ \bibinfo {author} {\bibfnamefont {M.}~\bibnamefont {Rovere}},\ }\href
  {\doibase 10.1103/PhysRevE.80.061502} {\bibfield  {journal} {\bibinfo
  {journal} {Phys. Rev. E}\ }\textbf {\bibinfo {volume} {80}},\ \bibinfo
  {pages} {061502} (\bibinfo {year} {2009})}\BibitemShut {NoStop}%
\bibitem [{\citenamefont {Gallo}\ \emph {et~al.}(2012)\citenamefont {Gallo},
  \citenamefont {Rovere},\ and\ \citenamefont {Chen}}]{Gallo:2012}%
  \BibitemOpen
  \bibfield  {author} {\bibinfo {author} {\bibfnamefont {P.}~\bibnamefont
  {Gallo}}, \bibinfo {author} {\bibfnamefont {M.}~\bibnamefont {Rovere}}, \
  and\ \bibinfo {author} {\bibfnamefont {S.-H.}\ \bibnamefont {Chen}},\ }\href
  {http://stacks.iop.org/0953-8984/24/i=6/a=064109} {\bibfield  {journal}
  {\bibinfo  {journal} {J. Phys.: Condens. Matter}\ }\textbf {\bibinfo {volume}
  {24}},\ \bibinfo {pages} {064109} (\bibinfo {year} {2012})}\BibitemShut
  {NoStop}%
\bibitem [{\citenamefont {Edmond}\ \emph {et~al.}(2012)\citenamefont {Edmond},
  \citenamefont {Nugent},\ and\ \citenamefont {Weeks}}]{Edmond:2012}%
  \BibitemOpen
  \bibfield  {author} {\bibinfo {author} {\bibfnamefont {K.~V.}\ \bibnamefont
  {Edmond}}, \bibinfo {author} {\bibfnamefont {C.~R.}\ \bibnamefont {Nugent}},
  \ and\ \bibinfo {author} {\bibfnamefont {E.~R.}\ \bibnamefont {Weeks}},\
  }\href {\doibase 10.1103/PhysRevE.85.041401} {\bibfield  {journal} {\bibinfo
  {journal} {Phys. Rev. E}\ }\textbf {\bibinfo {volume} {85}},\ \bibinfo
  {pages} {041401} (\bibinfo {year} {2012})}\BibitemShut {NoStop}%
\bibitem [{\citenamefont {Eral}\ \emph {et~al.}(2009)\citenamefont {Eral},
  \citenamefont {van~den Ende}, \citenamefont {Mugele},\ and\ \citenamefont
  {Duits}}]{Eral:2009}%
  \BibitemOpen
  \bibfield  {author} {\bibinfo {author} {\bibfnamefont {H.~B.}\ \bibnamefont
  {Eral}}, \bibinfo {author} {\bibfnamefont {D.}~\bibnamefont {van~den Ende}},
  \bibinfo {author} {\bibfnamefont {F.}~\bibnamefont {Mugele}}, \ and\ \bibinfo
  {author} {\bibfnamefont {M.~H.~G.}\ \bibnamefont {Duits}},\ }\href {\doibase
  10.1103/PhysRevE.80.061403} {\bibfield  {journal} {\bibinfo  {journal} {Phys.
  Rev. E}\ }\textbf {\bibinfo {volume} {80}},\ \bibinfo {pages} {061403}
  (\bibinfo {year} {2009})}\BibitemShut {NoStop}%
\bibitem [{\citenamefont {Ingebrigtsen}\ \emph {et~al.}(2013)\citenamefont
  {Ingebrigtsen}, \citenamefont {Errington}, \citenamefont {Truskett},\ and\
  \citenamefont {Dyre}}]{Ingebrigtsen:2013}%
  \BibitemOpen
  \bibfield  {author} {\bibinfo {author} {\bibfnamefont {T.~S.}\ \bibnamefont
  {Ingebrigtsen}}, \bibinfo {author} {\bibfnamefont {J.~R.}\ \bibnamefont
  {Errington}}, \bibinfo {author} {\bibfnamefont {T.~M.}\ \bibnamefont
  {Truskett}}, \ and\ \bibinfo {author} {\bibfnamefont {J.~C.}\ \bibnamefont
  {Dyre}},\ }\href {\doibase 10.1103/PhysRevLett.111.235901} {\bibfield
  {journal} {\bibinfo  {journal} {Phys. Rev. Lett.}\ }\textbf {\bibinfo
  {volume} {111}},\ \bibinfo {pages} {235901} (\bibinfo {year}
  {2013})}\BibitemShut {NoStop}%
\bibitem [{\citenamefont {Scheidler}\ \emph
  {et~al.}(2000{\natexlab{a}})\citenamefont {Scheidler}, \citenamefont {Kob},\
  and\ \citenamefont {Binder}}]{Scheidler:2000b}%
  \BibitemOpen
  \bibfield  {author} {\bibinfo {author} {\bibfnamefont {P.}~\bibnamefont
  {Scheidler}}, \bibinfo {author} {\bibfnamefont {W.}~\bibnamefont {Kob}}, \
  and\ \bibinfo {author} {\bibfnamefont {K.}~\bibnamefont {Binder}},\ }\href
  {http://stacks.iop.org/0295-5075/52/i=3/a=277} {\bibfield  {journal}
  {\bibinfo  {journal} {Europhys. Lett.}\ }\textbf {\bibinfo {volume} {52}},\
  \bibinfo {pages} {277} (\bibinfo {year} {2000}{\natexlab{a}})}\BibitemShut
  {NoStop}%
\bibitem [{\citenamefont {Scheidler}\ \emph
  {et~al.}(2000{\natexlab{b}})\citenamefont {Scheidler}, \citenamefont {Kob},\
  and\ \citenamefont {Binder}}]{Scheidler:2000a}%
  \BibitemOpen
  \bibfield  {author} {\bibinfo {author} {\bibfnamefont {P.}~\bibnamefont
  {Scheidler}}, \bibinfo {author} {\bibfnamefont {W.}~\bibnamefont {Kob}}, \
  and\ \bibinfo {author} {\bibfnamefont {K.}~\bibnamefont {Binder}},\ }\href
  {\doibase 10.1051/jp4:2000706} {\bibfield  {journal} {\bibinfo  {journal} {J.
  Phys. IV France}\ }\textbf {\bibinfo {volume} {10}},\ \bibinfo {pages} {Pr7-33}
  (\bibinfo {year} {2000}{\natexlab{b}})}\BibitemShut {NoStop}%
\bibitem [{\citenamefont {Scheidler}\ \emph {et~al.}(2002)\citenamefont
  {Scheidler}, \citenamefont {Kob},\ and\ \citenamefont
  {Binder}}]{Scheidler:2002}%
  \BibitemOpen
  \bibfield  {author} {\bibinfo {author} {\bibfnamefont {P.}~\bibnamefont
  {Scheidler}}, \bibinfo {author} {\bibfnamefont {W.}~\bibnamefont {Kob}}, \
  and\ \bibinfo {author} {\bibfnamefont {K.}~\bibnamefont {Binder}},\ }\href
  {http://stacks.iop.org/0295-5075/59/i=5/a=701} {\bibfield  {journal}
  {\bibinfo  {journal} {Europhys. Lett.}\ }\textbf {\bibinfo {volume} {59}},\
  \bibinfo {pages} {701} (\bibinfo {year} {2002})}\BibitemShut {NoStop}%
\bibitem [{\citenamefont {Evans}(1990)}]{Evans:1990}%
  \BibitemOpen
  \bibfield  {author} {\bibinfo {author} {\bibfnamefont {R.}~\bibnamefont
  {Evans}},\ }\href {http://stacks.iop.org/0953-8984/2/i=46/a=001} {\bibfield
  {journal} {\bibinfo  {journal} {J. Phys.: Condens. Matter}\ }\textbf
  {\bibinfo {volume} {2}},\ \bibinfo {pages} {8989} (\bibinfo {year}
  {1990})}\BibitemShut {NoStop}%
\bibitem [{\citenamefont {Antonchenko}\ \emph {et~al.}(1984)\citenamefont
  {Antonchenko}, \citenamefont {Ilyin}, \citenamefont {Makovsky}, \citenamefont
  {Pavlov},\ and\ \citenamefont {Sokhan}}]{Antonchenko:1984}%
  \BibitemOpen
  \bibfield  {author} {\bibinfo {author} {\bibfnamefont {V.}~\bibnamefont
  {Antonchenko}}, \bibinfo {author} {\bibfnamefont {V.}~\bibnamefont {Ilyin}},
  \bibinfo {author} {\bibfnamefont {N.}~\bibnamefont {Makovsky}}, \bibinfo
  {author} {\bibfnamefont {A.}~\bibnamefont {Pavlov}}, \ and\ \bibinfo {author}
  {\bibfnamefont {V.}~\bibnamefont {Sokhan}},\ }\href {\doibase
  10.1080/00268978400101261} {\bibfield  {journal} {\bibinfo  {journal} {Mol.
  Phys.}\ }\textbf {\bibinfo {volume} {52}},\ \bibinfo {pages} {345} (\bibinfo
  {year} {1984})}\BibitemShut {NoStop}%
\bibitem [{\citenamefont {Nyg\aa{}rd}\ \emph {et~al.}(2012)\citenamefont
  {Nyg\aa{}rd}, \citenamefont {Kjellander}, \citenamefont {Sarman},
  \citenamefont {Chodankar}, \citenamefont {Perret}, \citenamefont
  {Buitenhuis},\ and\ \citenamefont {van~der Veen}}]{Nygard:2012}%
  \BibitemOpen
  \bibfield  {author} {\bibinfo {author} {\bibfnamefont {K.}~\bibnamefont
  {Nyg\aa{}rd}}, \bibinfo {author} {\bibfnamefont {R.}~\bibnamefont
  {Kjellander}}, \bibinfo {author} {\bibfnamefont {S.}~\bibnamefont {Sarman}},
  \bibinfo {author} {\bibfnamefont {S.}~\bibnamefont {Chodankar}}, \bibinfo
  {author} {\bibfnamefont {E.}~\bibnamefont {Perret}}, \bibinfo {author}
  {\bibfnamefont {J.}~\bibnamefont {Buitenhuis}}, \ and\ \bibinfo {author}
  {\bibfnamefont {J.~F.}\ \bibnamefont {van~der Veen}},\ }\href {\doibase
  10.1103/PhysRevLett.108.037802} {\bibfield  {journal} {\bibinfo  {journal}
  {Phys. Rev. Lett.}\ }\textbf {\bibinfo {volume} {108}},\ \bibinfo {pages}
  {037802} (\bibinfo {year} {2012})}\BibitemShut {NoStop}%
\bibitem [{\citenamefont {Nyg\aa{}rd}\ \emph {et~al.}(2013)\citenamefont
  {Nyg\aa{}rd}, \citenamefont {Sarman},\ and\ \citenamefont
  {Kjellander}}]{Nygard:2013}%
  \BibitemOpen
  \bibfield  {author} {\bibinfo {author} {\bibfnamefont {K.}~\bibnamefont
  {Nyg\aa{}rd}}, \bibinfo {author} {\bibfnamefont {S.}~\bibnamefont {Sarman}},
  \ and\ \bibinfo {author} {\bibfnamefont {R.}~\bibnamefont {Kjellander}},\
  }\href {\doibase http://dx.doi.org/10.1063/1.4825176} {\bibfield  {journal}
  {\bibinfo  {journal} {J. Chem. Phys.}\ }\textbf {\bibinfo {volume} {139}},\
  \bibinfo {eid} {164701} (\bibinfo {year} {2013})}\BibitemShut {NoStop}%
\bibitem [{\citenamefont {Qi}\ \emph {et~al.}(2014)\citenamefont {Qi},
  \citenamefont {Gantapara},\ and\ \citenamefont {Dijkstra}}]{Qi:2014}%
  \BibitemOpen
  \bibfield  {author} {\bibinfo {author} {\bibfnamefont {W.}~\bibnamefont
  {Qi}}, \bibinfo {author} {\bibfnamefont {A.~P.}\ \bibnamefont {Gantapara}}, \
  and\ \bibinfo {author} {\bibfnamefont {M.}~\bibnamefont {Dijkstra}},\ }\href
  {\doibase 10.1039/C4SM00125G} {\bibfield  {journal} {\bibinfo  {journal}
  {Soft Matter}\ }\textbf {\bibinfo {volume} {10}},\ \bibinfo {pages} {5449}
  (\bibinfo {year} {2014})}\BibitemShut {NoStop}%
\bibitem [{\citenamefont {Strandburg}(1988)}]{Strandburg:1988}%
  \BibitemOpen
  \bibfield  {author} {\bibinfo {author} {\bibfnamefont {K.~J.}\ \bibnamefont
  {Strandburg}},\ }\href {\doibase 10.1103/RevModPhys.60.161} {\bibfield
  {journal} {\bibinfo  {journal} {Rev. Mod. Phys.}\ }\textbf {\bibinfo {volume}
  {60}},\ \bibinfo {pages} {161} (\bibinfo {year} {1988})}\BibitemShut
  {NoStop}%
\bibitem [{\citenamefont {Kosterlitz}\ and\ \citenamefont
  {Thouless}(1973)}]{Kosterlitz:1973}%
  \BibitemOpen
  \bibfield  {author} {\bibinfo {author} {\bibfnamefont {J.~M.}\ \bibnamefont
  {Kosterlitz}}\ and\ \bibinfo {author} {\bibfnamefont {D.~J.}\ \bibnamefont
  {Thouless}},\ }\href {http://stacks.iop.org/0022-3719/6/i=7/a=010} {\bibfield
   {journal} {\bibinfo  {journal} {J. Phys. C: Solid State Physics}\ }\textbf
  {\bibinfo {volume} {6}},\ \bibinfo {pages} {1181} (\bibinfo {year}
  {1973})}\BibitemShut {NoStop}%
\bibitem [{\citenamefont {Young}(1979)}]{Young:1979}%
  \BibitemOpen
  \bibfield  {author} {\bibinfo {author} {\bibfnamefont {A.~P.}\ \bibnamefont
  {Young}},\ }\href {\doibase 10.1103/PhysRevB.19.1855} {\bibfield  {journal}
  {\bibinfo  {journal} {Phys. Rev. B}\ }\textbf {\bibinfo {volume} {19}},\
  \bibinfo {pages} {1855} (\bibinfo {year} {1979})}\BibitemShut {NoStop}%
\bibitem [{\citenamefont {Nelson}\ and\ \citenamefont
  {Halperin}(1979)}]{Nelson:1979}%
  \BibitemOpen
  \bibfield  {author} {\bibinfo {author} {\bibfnamefont {D.~R.}\ \bibnamefont
  {Nelson}}\ and\ \bibinfo {author} {\bibfnamefont {B.~I.}\ \bibnamefont
  {Halperin}},\ }\href {\doibase 10.1103/PhysRevB.19.2457} {\bibfield
  {journal} {\bibinfo  {journal} {Phys. Rev. B}\ }\textbf {\bibinfo {volume}
  {19}},\ \bibinfo {pages} {2457} (\bibinfo {year} {1979})}\BibitemShut
  {NoStop}%
\bibitem [{\citenamefont {Bernard}\ and\ \citenamefont
  {Krauth}(2011)}]{Bernard:2011}%
  \BibitemOpen
  \bibfield  {author} {\bibinfo {author} {\bibfnamefont {E.~P.}\ \bibnamefont
  {Bernard}}\ and\ \bibinfo {author} {\bibfnamefont {W.}~\bibnamefont
  {Krauth}},\ }\href {\doibase 10.1103/PhysRevLett.107.155704} {\bibfield
  {journal} {\bibinfo  {journal} {Phys. Rev. Lett.}\ }\textbf {\bibinfo
  {volume} {107}},\ \bibinfo {pages} {155704} (\bibinfo {year}
  {2011})}\BibitemShut {NoStop}%
\bibitem [{\citenamefont {Kapfer}\ and\ \citenamefont
  {Krauth}(2014)}]{Kapfer:2014}%
  \BibitemOpen
  \bibfield  {author} {\bibinfo {author} {\bibfnamefont {S.}~\bibnamefont
  {Kapfer}}\ and\ \bibinfo {author} {\bibfnamefont {W.}~\bibnamefont
  {Krauth}},\ }\href@noop {} {\bibfield  {journal} {\bibinfo  {journal}
  {arXiv:1406.7224}\ } (\bibinfo {year} {2014})}\BibitemShut {NoStop}%
\bibitem [{\citenamefont {Bengtzelius}\ \emph {et~al.}(1984)\citenamefont
  {Bengtzelius}, \citenamefont {G\"otze},\ and\ \citenamefont
  {Sj\"olander}}]{Bengtzelius:1984}%
  \BibitemOpen
  \bibfield  {author} {\bibinfo {author} {\bibfnamefont {U.}~\bibnamefont
  {Bengtzelius}}, \bibinfo {author} {\bibfnamefont {W.}~\bibnamefont
  {G\"otze}}, \ and\ \bibinfo {author} {\bibfnamefont {A.}~\bibnamefont
  {Sj\"olander}},\ }\href {http://stacks.iop.org/0022-3719/17/i=33/a=005}
  {\bibfield  {journal} {\bibinfo  {journal} {J. Phys. C: Solid State Phys.}\
  }\textbf {\bibinfo {volume} {17}},\ \bibinfo {pages} {5915} (\bibinfo {year}
  {1984})}\BibitemShut {NoStop}%
\bibitem [{\citenamefont {G\"otze}(2009)}]{Goetze:Complex_Dynamics}%
  \BibitemOpen
  \bibfield  {author} {\bibinfo {author} {\bibfnamefont {W.}~\bibnamefont
  {G\"otze}},\ }\href@noop {} {\emph {\bibinfo {title} {Complex Dynamics of
  Glass-Forming Liquids -- A Mode-Coupling Theory}}}\ (\bibinfo  {publisher}
  {Oxford University Press},\ \bibinfo {address} {Oxford},\ \bibinfo {year} {2009})\BibitemShut
  {NoStop}%
\bibitem [{\citenamefont {G\"otze}(1999)}]{Goetze:1999}%
  \BibitemOpen
  \bibfield  {author} {\bibinfo {author} {\bibfnamefont {W.}~\bibnamefont
  {G\"otze}},\ }\href {http://stacks.iop.org/0953-8984/11/i=10A/a=002}
  {\bibfield  {journal} {\bibinfo  {journal} {J. Phys.: Condens. Matter}\
  }\textbf {\bibinfo {volume} {11}},\ \bibinfo {pages} {A1} (\bibinfo {year}
  {1999})}\BibitemShut {NoStop}%
\bibitem [{\citenamefont {Bayer}\ \emph {et~al.}(2007)\citenamefont {Bayer},
  \citenamefont {Brader}, \citenamefont {Ebert}, \citenamefont {Fuchs},
  \citenamefont {Lange}, \citenamefont {Maret}, \citenamefont {Schilling},
  \citenamefont {Sperl},\ and\ \citenamefont {Wittmer}}]{Bayer:2007}%
  \BibitemOpen
  \bibfield  {author} {\bibinfo {author} {\bibfnamefont {M.}~\bibnamefont
  {Bayer}}, \bibinfo {author} {\bibfnamefont {J.~M.}\ \bibnamefont {Brader}},
  \bibinfo {author} {\bibfnamefont {F.}~\bibnamefont {Ebert}}, \bibinfo
  {author} {\bibfnamefont {M.}~\bibnamefont {Fuchs}}, \bibinfo {author}
  {\bibfnamefont {E.}~\bibnamefont {Lange}}, \bibinfo {author} {\bibfnamefont
  {G.}~\bibnamefont {Maret}}, \bibinfo {author} {\bibfnamefont
  {R.}~\bibnamefont {Schilling}}, \bibinfo {author} {\bibfnamefont
  {M.}~\bibnamefont {Sperl}}, \ and\ \bibinfo {author} {\bibfnamefont {J.~P.}\
  \bibnamefont {Wittmer}},\ }\href {\doibase 10.1103/PhysRevE.76.011508}
  {\bibfield  {journal} {\bibinfo  {journal} {Phys. Rev. E}\ }\textbf {\bibinfo
  {volume} {76}},\ \bibinfo {pages} {011508} (\bibinfo {year}
  {2007})}\BibitemShut {NoStop}%
\bibitem [{\citenamefont {Hajnal}\ \emph {et~al.}(2009)\citenamefont {Hajnal},
  \citenamefont {Brader},\ and\ \citenamefont {Schilling}}]{Hajnal:2009}%
  \BibitemOpen
  \bibfield  {author} {\bibinfo {author} {\bibfnamefont {D.}~\bibnamefont
  {Hajnal}}, \bibinfo {author} {\bibfnamefont {J.~M.}\ \bibnamefont {Brader}},
  \ and\ \bibinfo {author} {\bibfnamefont {R.}~\bibnamefont {Schilling}},\
  }\href {\doibase 10.1103/PhysRevE.80.021503} {\bibfield  {journal} {\bibinfo
  {journal} {Phys. Rev. E}\ }\textbf {\bibinfo {volume} {80}},\ \bibinfo
  {pages} {021503} (\bibinfo {year} {2009})}\BibitemShut {NoStop}%
\bibitem [{\citenamefont {Hajnal}\ \emph {et~al.}(2011)\citenamefont {Hajnal},
  \citenamefont {Oettel},\ and\ \citenamefont {Schilling}}]{Hajnal:2011}%
  \BibitemOpen
  \bibfield  {author} {\bibinfo {author} {\bibfnamefont {D.}~\bibnamefont
  {Hajnal}}, \bibinfo {author} {\bibfnamefont {M.}~\bibnamefont {Oettel}}, \
  and\ \bibinfo {author} {\bibfnamefont {R.}~\bibnamefont {Schilling}},\ }\href
  {\doibase 10.1016/j.jnoncrysol.2010.06.039} {\bibfield  {journal} {\bibinfo
  {journal} {J. Non-Cryst. Solids}\ }\textbf {\bibinfo {volume} {357}},\
  \bibinfo {pages} {302 } (\bibinfo {year} {2011})}\BibitemShut {NoStop}%
\bibitem [{\citenamefont {Weysser}\ and\ \citenamefont
  {Hajnal}(2011)}]{Weysser:2011}%
  \BibitemOpen
  \bibfield  {author} {\bibinfo {author} {\bibfnamefont {F.}~\bibnamefont
  {Weysser}}\ and\ \bibinfo {author} {\bibfnamefont {D.}~\bibnamefont
  {Hajnal}},\ }\href {\doibase 10.1103/PhysRevE.83.041503} {\bibfield
  {journal} {\bibinfo  {journal} {Phys. Rev. E}\ }\textbf {\bibinfo {volume}
  {83}},\ \bibinfo {pages} {041503} (\bibinfo {year} {2011})}\BibitemShut
  {NoStop}%
\bibitem [{\citenamefont {{H. K\"onig}}\ \emph {et~al.}(2005)\citenamefont {{H.
  K\"onig}}, \citenamefont {{R. Hund}}, \citenamefont {{K. Zahn}},\ and\
  \citenamefont {{G. Maret}}}]{Koenig:2005}%
  \BibitemOpen
  \bibfield  {author} {\bibinfo {author} {\bibnamefont {{H. K\"onig}}},
  \bibinfo {author} {\bibnamefont {{R. Hund}}}, \bibinfo {author} {\bibnamefont
  {{K. Zahn}}}, \ and\ \bibinfo {author} {\bibnamefont {{G. Maret}}},\ }\href
  {\doibase 10.1140/epje/e2005-00034-9} {\bibfield  {journal} {\bibinfo
  {journal} {Eur. Phys. J. E}\ }\textbf {\bibinfo {volume} {18}},\ \bibinfo
  {pages} {287} (\bibinfo {year} {2005})}\BibitemShut {NoStop}%
\bibitem [{\citenamefont {Lang}\ \emph {et~al.}(2010)\citenamefont {Lang},
  \citenamefont {Bo\ifmmode~\mbox{\c{t}}\else \c{t}\fi{}an}, \citenamefont
  {Oettel}, \citenamefont {Hajnal}, \citenamefont {Franosch},\ and\
  \citenamefont {Schilling}}]{Lang:2010}%
  \BibitemOpen
  \bibfield  {author} {\bibinfo {author} {\bibfnamefont {S.}~\bibnamefont
  {Lang}}, \bibinfo {author} {\bibfnamefont {V.}~\bibnamefont
  {Bo\ifmmode~\mbox{\c{t}}\else \c{t}\fi{}an}}, \bibinfo {author}
  {\bibfnamefont {M.}~\bibnamefont {Oettel}}, \bibinfo {author} {\bibfnamefont
  {D.}~\bibnamefont {Hajnal}}, \bibinfo {author} {\bibfnamefont
  {T.}~\bibnamefont {Franosch}}, \ and\ \bibinfo {author} {\bibfnamefont
  {R.}~\bibnamefont {Schilling}},\ }\href {\doibase
  10.1103/PhysRevLett.105.125701} {\bibfield  {journal} {\bibinfo  {journal}
  {Phys. Rev. Lett.}\ }\textbf {\bibinfo {volume} {105}},\ \bibinfo {pages}
  {125701} (\bibinfo {year} {2010})}\BibitemShut {NoStop}%
\bibitem [{\citenamefont {Lang}\ \emph {et~al.}(2012)\citenamefont {Lang},
  \citenamefont {Schilling}, \citenamefont {Krakoviack},\ and\ \citenamefont
  {Franosch}}]{Lang:2012}%
  \BibitemOpen
  \bibfield  {author} {\bibinfo {author} {\bibfnamefont {S.}~\bibnamefont
  {Lang}}, \bibinfo {author} {\bibfnamefont {R.}~\bibnamefont {Schilling}},
  \bibinfo {author} {\bibfnamefont {V.}~\bibnamefont {Krakoviack}}, \ and\
  \bibinfo {author} {\bibfnamefont {T.}~\bibnamefont {Franosch}},\ }\href
  {\doibase 10.1103/PhysRevE.86.021502} {\bibfield  {journal} {\bibinfo
  {journal} {Phys. Rev. E}\ }\textbf {\bibinfo {volume} {86}},\ \bibinfo
  {pages} {021502} (\bibinfo {year} {2012})}\BibitemShut {NoStop}%
\bibitem [{\citenamefont {Lang}\ and\ \citenamefont
  {Franosch}(2014)}]{Lang:2014b}%
  \BibitemOpen
  \bibfield  {author} {\bibinfo {author} {\bibfnamefont {S.}~\bibnamefont
  {Lang}}\ and\ \bibinfo {author} {\bibfnamefont {T.}~\bibnamefont
  {Franosch}},\ }\href {\doibase 10.1103/PhysRevE.89.062122} {\bibfield
  {journal} {\bibinfo  {journal} {Phys. Rev. E}\ }\textbf {\bibinfo {volume}
  {89}},\ \bibinfo {pages} {062122} (\bibinfo {year} {2014})}\BibitemShut
  {NoStop}%
\bibitem [{\citenamefont {Krakoviack}(2005)}]{Krakoviack:2005}%
  \BibitemOpen
  \bibfield  {author} {\bibinfo {author} {\bibfnamefont {V.}~\bibnamefont
  {Krakoviack}},\ }\href {\doibase 10.1103/PhysRevLett.94.065703} {\bibfield
  {journal} {\bibinfo  {journal} {Phys. Rev. Lett.}\ }\textbf {\bibinfo
  {volume} {94}},\ \bibinfo {pages} {065703} (\bibinfo {year}
  {2005})}\BibitemShut {NoStop}%
\bibitem [{\citenamefont {Krakoviack}(2007)}]{Krakoviack:2007}%
  \BibitemOpen
  \bibfield  {author} {\bibinfo {author} {\bibfnamefont {V.}~\bibnamefont
  {Krakoviack}},\ }\href {\doibase 10.1103/PhysRevE.75.031503} {\bibfield
  {journal} {\bibinfo  {journal} {Phys. Rev. E}\ }\textbf {\bibinfo {volume}
  {75}},\ \bibinfo {eid} {031503} (\bibinfo {year} {2007})}\BibitemShut
  {NoStop}%
\bibitem [{\citenamefont {Krakoviack}(2009)}]{Krakoviack:2009}%
  \BibitemOpen
  \bibfield  {author} {\bibinfo {author} {\bibfnamefont {V.}~\bibnamefont
  {Krakoviack}},\ }\href {\doibase 10.1103/PhysRevE.79.061501} {\bibfield
  {journal} {\bibinfo  {journal} {Phys. Rev. E}\ }\textbf {\bibinfo {volume}
  {79}},\ \bibinfo {eid} {061501} (\bibinfo {year} {2009})}\BibitemShut
  {NoStop}%
\bibitem [{\citenamefont {Krakoviack}(2011)}]{Krakoviack:2011}%
  \BibitemOpen
  \bibfield  {author} {\bibinfo {author} {\bibfnamefont {V.}~\bibnamefont
  {Krakoviack}},\ }\href {\doibase 10.1103/PhysRevE.84.050501} {\bibfield
  {journal} {\bibinfo  {journal} {Phys. Rev. E}\ }\textbf {\bibinfo {volume}
  {84}},\ \bibinfo {pages} {050501} (\bibinfo {year} {2011})}\BibitemShut
  {NoStop}%
\bibitem [{\citenamefont {Szamel}\ and\ \citenamefont
  {Flenner}(2013)}]{Szamel:2013}%
  \BibitemOpen
  \bibfield  {author} {\bibinfo {author} {\bibfnamefont {G.}~\bibnamefont
  {Szamel}}\ and\ \bibinfo {author} {\bibfnamefont {E.}~\bibnamefont
  {Flenner}},\ }\href {http://stacks.iop.org/0295-5075/101/i=6/a=66005}
  {\bibfield  {journal} {\bibinfo  {journal} {EPL (Europhysics Letters)}\
  }\textbf {\bibinfo {volume} {101}},\ \bibinfo {pages} {66005} (\bibinfo
  {year} {2013})}\BibitemShut {NoStop}%
\bibitem [{\citenamefont {Mandal}\ \emph {et~al.}(2014)\citenamefont {Mandal},
  \citenamefont {Lang}, \citenamefont {Gross}, \citenamefont {Oettel},
  \citenamefont {Raabe}, \citenamefont {Franosch},\ and\ \citenamefont
  {Varnik}}]{Mandal:2014}%
  \BibitemOpen
  \bibfield  {author} {\bibinfo {author} {\bibfnamefont {S.}~\bibnamefont
  {Mandal}}, \bibinfo {author} {\bibfnamefont {S.}~\bibnamefont {Lang}},
  \bibinfo {author} {\bibfnamefont {M.}~\bibnamefont {Gross}}, \bibinfo
  {author} {\bibfnamefont {M.}~\bibnamefont {Oettel}}, \bibinfo {author}
  {\bibfnamefont {D.}~\bibnamefont {Raabe}}, \bibinfo {author} {\bibfnamefont
  {T.}~\bibnamefont {Franosch}}, \ and\ \bibinfo {author} {\bibfnamefont
  {F.}~\bibnamefont {Varnik}},\ }\href {\doibase 10.1038/ncomms5435} {\bibfield
   {journal} {\bibinfo  {journal} {Nature Communications}\ }\textbf {\bibinfo
  {volume} {5}},\ \bibinfo {pages} {4435} (\bibinfo {year} {2014})}\BibitemShut
  {NoStop}%
\bibitem [{\citenamefont {Lang}\ \emph {et~al.}(2014)\citenamefont {Lang},
  \citenamefont {Franosch},\ and\ \citenamefont {Schilling}}]{Lang:2014a}%
  \BibitemOpen
  \bibfield  {author} {\bibinfo {author} {\bibfnamefont {S.}~\bibnamefont
  {Lang}}, \bibinfo {author} {\bibfnamefont {T.}~\bibnamefont {Franosch}}, \
  and\ \bibinfo {author} {\bibfnamefont {R.}~\bibnamefont {Schilling}},\ }\href
  {\doibase http://dx.doi.org/10.1063/1.4867284} {\bibfield  {journal}
  {\bibinfo  {journal} {J. Chem. Phys}\ }\textbf {\bibinfo {volume} {140}},\
  \bibinfo {eid} {104506} (\bibinfo {year} {2014})}\BibitemShut {NoStop}%
\bibitem [{\citenamefont {Schilling}\ and\ \citenamefont
  {Scheidsteger}(1997)}]{Scheidsteger:1997}%
  \BibitemOpen
  \bibfield  {author} {\bibinfo {author} {\bibfnamefont {R.}~\bibnamefont
  {Schilling}}\ and\ \bibinfo {author} {\bibfnamefont {T.}~\bibnamefont
  {Scheidsteger}},\ }\href {\doibase 10.1103/PhysRevE.56.2932} {\bibfield
  {journal} {\bibinfo  {journal} {Phys. Rev. E}\ }\textbf {\bibinfo {volume}
  {56}},\ \bibinfo {pages} {2932} (\bibinfo {year} {1997})}\BibitemShut
  {NoStop}%
\bibitem [{\citenamefont {Franosch}\ \emph {et~al.}(1997)\citenamefont
  {Franosch}, \citenamefont {Fuchs}, \citenamefont {G{\"o}tze}, \citenamefont
  {Mayr},\ and\ \citenamefont {Singh}}]{Franosch:1997}%
  \BibitemOpen
  \bibfield  {author} {\bibinfo {author} {\bibfnamefont {T.}~\bibnamefont
  {Franosch}}, \bibinfo {author} {\bibfnamefont {M.}~\bibnamefont {Fuchs}},
  \bibinfo {author} {\bibfnamefont {W.}~\bibnamefont {G{\"o}tze}}, \bibinfo
  {author} {\bibfnamefont {M.~R.}\ \bibnamefont {Mayr}}, \ and\ \bibinfo
  {author} {\bibfnamefont {A.~P.}\ \bibnamefont {Singh}},\ }\href {\doibase
  10.1103/PhysRevE.56.5659} {\bibfield  {journal} {\bibinfo  {journal} {{Phys.
  Rev. E}}\ }\textbf {\bibinfo {volume} {{56}}},\ \bibinfo {pages} {5659}
  (\bibinfo {year} {1997})}\BibitemShut {NoStop}%
\bibitem [{\citenamefont {Forster}(1975)}]{Forster:Hydrodynamic_Fluctuations}%
  \BibitemOpen
  \bibfield  {author} {\bibinfo {author} {\bibfnamefont {D.}~\bibnamefont
  {Forster}},\ }\href@noop {} {\emph {\bibinfo {title} {Hydrodynamic
  Fluctuations, Broken Symmetry, And Correlation Functions}}}\ (\bibinfo
  {publisher} {Benjamin},\ \bibinfo {address}
  {San Francisco},\ \bibinfo {year} {1975})\BibitemShut {NoStop}%
\bibitem [{\citenamefont
  {Henderson}(1992)}]{Henderson:Fundamentals_of_inhomogeneous_fluids}%
  \BibitemOpen
  \bibfield  {author} {\bibinfo {author} {\bibfnamefont {D.}~\bibnamefont
  {Henderson}},\ }\href@noop {} {\emph {\bibinfo {title} {Fundamentals of
  inhomogeneous fluids}}}\ (\bibinfo  {publisher} {Dekker},\ \bibinfo {address}
  {New York},\ \bibinfo {year} {1992})\BibitemShut {NoStop}%
\bibitem [{\citenamefont {G{\"o}tze}\ and\ \citenamefont
  {Sj\"ogren}(1995)}]{Goetze:1995}%
  \BibitemOpen
  \bibfield  {author} {\bibinfo {author} {\bibfnamefont {W.}~\bibnamefont
  {G{\"o}tze}}\ and\ \bibinfo {author} {\bibfnamefont {L.}~\bibnamefont
  {Sj\"ogren}},\ }\href {\doibase 10.1006/jmaa.1995.1352} {\bibfield  {journal}
  {\bibinfo  {journal} {J. Math. Anal. Appl.}\ }\textbf {\bibinfo {volume}
  {195}},\ \bibinfo {pages} {230} (\bibinfo {year} {1995})}\BibitemShut
  {NoStop}%
\bibitem [{\citenamefont {Franosch}\ and\ \citenamefont
  {Voigtmann}(2002)}]{Franosch:2002}%
  \BibitemOpen
  \bibfield  {author} {\bibinfo {author} {\bibfnamefont {T.}~\bibnamefont
  {Franosch}}\ and\ \bibinfo {author} {\bibfnamefont {{\relax
  Th}.}~\bibnamefont {Voigtmann}},\ }\href {\doibase 10.1023/A:1019991729106}
  {\bibfield  {journal} {\bibinfo  {journal} {{J. Stat. Phys.}}\ }\textbf
  {\bibinfo {volume} {{109}}},\ \bibinfo {pages} {{237}} (\bibinfo {year}
  {2002})}\BibitemShut {NoStop}%
\bibitem [{\citenamefont {Lang}\ \emph {et~al.}(2013)\citenamefont {Lang},
  \citenamefont {Schilling},\ and\ \citenamefont {Franosch}}]{Lang:2013}%
  \BibitemOpen
  \bibfield  {author} {\bibinfo {author} {\bibfnamefont {S.}~\bibnamefont
  {Lang}}, \bibinfo {author} {\bibfnamefont {R.}~\bibnamefont {Schilling}}, \
  and\ \bibinfo {author} {\bibfnamefont {T.}~\bibnamefont {Franosch}},\ }\href
  {http://stacks.iop.org/1742-5468/2013/i=12/a=P12007} {\bibfield  {journal}
  {\bibinfo  {journal} {J. Stat. Mech.: Theory and Experiment}\ }\textbf
  {\bibinfo {volume} {2013}},\ \bibinfo {pages} {P12007} (\bibinfo {year}
  {2013})}\BibitemShut {NoStop}%
\bibitem [{\citenamefont {Franosch}\ \emph {et~al.}(2012)\citenamefont
  {Franosch}, \citenamefont {Lang},\ and\ \citenamefont
  {Schilling}}]{Franosch:2012}%
  \BibitemOpen
  \bibfield  {author} {\bibinfo {author} {\bibfnamefont {T.}~\bibnamefont
  {Franosch}}, \bibinfo {author} {\bibfnamefont {S.}~\bibnamefont {Lang}}, \
  and\ \bibinfo {author} {\bibfnamefont {R.}~\bibnamefont {Schilling}},\ }\href
  {\doibase 10.1103/PhysRevLett.109.240601} {\bibfield  {journal} {\bibinfo
  {journal} {Phys. Rev. Lett.}\ }\textbf {\bibinfo {volume} {109}},\ \bibinfo
  {pages} {240601} (\bibinfo {year} {2012})}\BibitemShut {NoStop}%
\bibitem [{\citenamefont {Schilling}(2002)}]{Schilling:2002}%
  \BibitemOpen
  \bibfield  {author} {\bibinfo {author} {\bibfnamefont {R.}~\bibnamefont
  {Schilling}},\ }\href {\doibase 10.1103/PhysRevE.65.051206} {\bibfield
  {journal} {\bibinfo  {journal} {Phys. Rev. E}\ }\textbf {\bibinfo {volume}
  {65}},\ \bibinfo {pages} {051206} (\bibinfo {year} {2002})}\BibitemShut
  {NoStop}%
\bibitem [{\citenamefont {van Megen}\ and\ \citenamefont
  {Underwood}(1994)}]{Megen:1994}%
  \BibitemOpen
  \bibfield  {author} {\bibinfo {author} {\bibfnamefont {W.}~\bibnamefont {van
  Megen}}\ and\ \bibinfo {author} {\bibfnamefont {S.~M.}\ \bibnamefont
  {Underwood}},\ }\href {\doibase 10.1103/PhysRevE.49.4206} {\bibfield
  {journal} {\bibinfo  {journal} {Phys. Rev. E}\ }\textbf {\bibinfo {volume}
  {49}},\ \bibinfo {pages} {4206} (\bibinfo {year} {1994})}\BibitemShut
  {NoStop}%
\bibitem [{\citenamefont {Kob}\ and\ \citenamefont
  {Andersen}(1994)}]{Kob:1994}%
  \BibitemOpen
  \bibfield  {author} {\bibinfo {author} {\bibfnamefont {W.}~\bibnamefont
  {Kob}}\ and\ \bibinfo {author} {\bibfnamefont {H.~C.}\ \bibnamefont
  {Andersen}},\ }\href {\doibase 10.1103/PhysRevLett.73.1376} {\bibfield
  {journal} {\bibinfo  {journal} {Phys. Rev. Lett.}\ }\textbf {\bibinfo
  {volume} {73}},\ \bibinfo {pages} {1376} (\bibinfo {year}
  {1994})}\BibitemShut {NoStop}%
\bibitem [{\citenamefont {Kob}\ and\ \citenamefont
  {Andersen}(1995{\natexlab{a}})}]{Kob:1995a}%
  \BibitemOpen
  \bibfield  {author} {\bibinfo {author} {\bibfnamefont {W.}~\bibnamefont
  {Kob}}\ and\ \bibinfo {author} {\bibfnamefont {H.~C.}\ \bibnamefont
  {Andersen}},\ }\href {\doibase 10.1103/PhysRevE.51.4626} {\bibfield
  {journal} {\bibinfo  {journal} {Phys. Rev. E}\ }\textbf {\bibinfo {volume}
  {51}},\ \bibinfo {pages} {4626} (\bibinfo {year}
  {1995}{\natexlab{a}})}\BibitemShut {NoStop}%
\bibitem [{\citenamefont {Kob}\ and\ \citenamefont
  {Andersen}(1995{\natexlab{b}})}]{Kob:1995b}%
  \BibitemOpen
  \bibfield  {author} {\bibinfo {author} {\bibfnamefont {W.}~\bibnamefont
  {Kob}}\ and\ \bibinfo {author} {\bibfnamefont {H.~C.}\ \bibnamefont
  {Andersen}},\ }\href {\doibase {10.1103/PhysRevE.52.4134}} {\bibfield
  {journal} {\bibinfo  {journal} {{Phys. Rev. E}}\ }\textbf {\bibinfo {volume}
  {{52}}},\ \bibinfo {pages} {4134} (\bibinfo {year}
  {{1995}}{\natexlab{b}})}\BibitemShut {NoStop}%
\bibitem [{\citenamefont {Hansen}\ and\ \citenamefont
  {McDonald}(2006)}]{Hansen:Theory_of_Simple_Liquids}%
  \BibitemOpen
  \bibfield  {author} {\bibinfo {author} {\bibfnamefont {J.~P.}\ \bibnamefont
  {Hansen}}\ and\ \bibinfo {author} {\bibfnamefont {I.~R.}\ \bibnamefont
  {McDonald}},\ }\href@noop {} {\emph {\bibinfo {title} {Theory of Simple
  Liquids}}}\ (\bibinfo  {publisher} {Academic Press},\ \bibinfo {address}
  {Waltham},\ \bibinfo {year}
  {2006})\BibitemShut {NoStop}%
\bibitem [{\citenamefont {Zaccarelli}\ \emph {et~al.}(2004)\citenamefont
  {Zaccarelli}, \citenamefont {L\"owen}, \citenamefont {Wessels}, \citenamefont
  {Sciortino}, \citenamefont {Tartaglia},\ and\ \citenamefont
  {Likos}}]{Zaccarelli:2004}%
  \BibitemOpen
  \bibfield  {author} {\bibinfo {author} {\bibfnamefont {E.}~\bibnamefont
  {Zaccarelli}}, \bibinfo {author} {\bibfnamefont {H.}~\bibnamefont {L\"owen}},
  \bibinfo {author} {\bibfnamefont {P.~P.~F.}\ \bibnamefont {Wessels}},
  \bibinfo {author} {\bibfnamefont {F.}~\bibnamefont {Sciortino}}, \bibinfo
  {author} {\bibfnamefont {P.}~\bibnamefont {Tartaglia}}, \ and\ \bibinfo
  {author} {\bibfnamefont {C.~N.}\ \bibnamefont {Likos}},\ }\href {\doibase
  10.1103/PhysRevLett.92.225703} {\bibfield  {journal} {\bibinfo  {journal}
  {Phys. Rev. Lett.}\ }\textbf {\bibinfo {volume} {92}},\ \bibinfo {pages}
  {225703} (\bibinfo {year} {2004})}\BibitemShut {NoStop}%
\bibitem [{\citenamefont {Pham}\ \emph {et~al.}(2002)\citenamefont {Pham},
  \citenamefont {Puertas}, \citenamefont {Bergenholtz}, \citenamefont
  {Egelhaaf}, \citenamefont {Moussa{\"i}d}, \citenamefont {Pusey}, \citenamefont
  {Schofield}, \citenamefont {Cates}, \citenamefont {Fuchs},\ and\
  \citenamefont {Poon}}]{Pham:2002}%
  \BibitemOpen
  \bibfield  {author} {\bibinfo {author} {\bibfnamefont {K.~N.}\ \bibnamefont
  {Pham}}, \bibinfo {author} {\bibfnamefont {A.~M.}\ \bibnamefont {Puertas}},
  \bibinfo {author} {\bibfnamefont {J.}~\bibnamefont {Bergenholtz}}, \bibinfo
  {author} {\bibfnamefont {S.~U.}\ \bibnamefont {Egelhaaf}}, \bibinfo {author}
  {\bibfnamefont {A.}~\bibnamefont {Moussa{\"i}d}}, \bibinfo {author}
  {\bibfnamefont {P.~N.}\ \bibnamefont {Pusey}}, \bibinfo {author}
  {\bibfnamefont {A.~B.}\ \bibnamefont {Schofield}}, \bibinfo {author}
  {\bibfnamefont {M.~E.}\ \bibnamefont {Cates}}, \bibinfo {author}
  {\bibfnamefont {M.}~\bibnamefont {Fuchs}}, \ and\ \bibinfo {author}
  {\bibfnamefont {W.~C.~K.}\ \bibnamefont {Poon}},\ }\href {\doibase
  10.1126/science.1068238} {\bibfield  {journal} {\bibinfo  {journal}
  {Science}\ }\textbf {\bibinfo {volume} {296}},\ \bibinfo {pages} {104}
  (\bibinfo {year} {2002})}\BibitemShut {NoStop}%
\bibitem [{\citenamefont {Foffi}\ \emph {et~al.}(2003)\citenamefont {Foffi},
  \citenamefont {Sciortino}, \citenamefont {Tartaglia}, \citenamefont
  {Zaccarelli}, \citenamefont {Lo~Verso}, \citenamefont {Reatto}, \citenamefont
  {Dawson},\ and\ \citenamefont {Likos}}]{Foffi:2003}%
  \BibitemOpen
  \bibfield  {author} {\bibinfo {author} {\bibfnamefont {G.}~\bibnamefont
  {Foffi}}, \bibinfo {author} {\bibfnamefont {F.}~\bibnamefont {Sciortino}},
  \bibinfo {author} {\bibfnamefont {P.}~\bibnamefont {Tartaglia}}, \bibinfo
  {author} {\bibfnamefont {E.}~\bibnamefont {Zaccarelli}}, \bibinfo {author}
  {\bibfnamefont {F.}~\bibnamefont {Lo~Verso}}, \bibinfo {author}
  {\bibfnamefont {L.}~\bibnamefont {Reatto}}, \bibinfo {author} {\bibfnamefont
  {K.~A.}\ \bibnamefont {Dawson}}, \ and\ \bibinfo {author} {\bibfnamefont
  {C.~N.}\ \bibnamefont {Likos}},\ }\href {\doibase
  10.1103/PhysRevLett.90.238301} {\bibfield  {journal} {\bibinfo  {journal}
  {Phys. Rev. Lett.}\ }\textbf {\bibinfo {volume} {90}},\ \bibinfo {pages}
  {238301} (\bibinfo {year} {2003})}\BibitemShut {NoStop}%
\bibitem [{\citenamefont {Dawson}\ \emph {et~al.}(2000)\citenamefont {Dawson},
  \citenamefont {Foffi}, \citenamefont {Fuchs}, \citenamefont {G\"otze},
  \citenamefont {Sciortino}, \citenamefont {Sperl}, \citenamefont {Tartaglia},
  \citenamefont {Voigtmann},\ and\ \citenamefont {Zaccarelli}}]{Dawson:2000}%
  \BibitemOpen
  \bibfield  {author} {\bibinfo {author} {\bibfnamefont {K.}~\bibnamefont
  {Dawson}}, \bibinfo {author} {\bibfnamefont {G.}~\bibnamefont {Foffi}},
  \bibinfo {author} {\bibfnamefont {M.}~\bibnamefont {Fuchs}}, \bibinfo
  {author} {\bibfnamefont {W.}~\bibnamefont {G\"otze}}, \bibinfo {author}
  {\bibfnamefont {F.}~\bibnamefont {Sciortino}}, \bibinfo {author}
  {\bibfnamefont {M.}~\bibnamefont {Sperl}}, \bibinfo {author} {\bibfnamefont
  {P.}~\bibnamefont {Tartaglia}}, \bibinfo {author} {\bibfnamefont {{\relax
  Th}.}~\bibnamefont {Voigtmann}}, \ and\ \bibinfo {author} {\bibfnamefont
  {E.}~\bibnamefont {Zaccarelli}},\ }\href {\doibase
  10.1103/PhysRevE.63.011401} {\bibfield  {journal} {\bibinfo  {journal} {Phys.
  Rev. E}\ }\textbf {\bibinfo {volume} {63}},\ \bibinfo {pages} {011401}
  (\bibinfo {year} {2000})}\BibitemShut {NoStop}%
\bibitem [{\citenamefont {Eckert}\ and\ \citenamefont
  {Bartsch}(2002)}]{Eckert:2002}%
  \BibitemOpen
  \bibfield  {author} {\bibinfo {author} {\bibfnamefont {T.}~\bibnamefont
  {Eckert}}\ and\ \bibinfo {author} {\bibfnamefont {E.}~\bibnamefont
  {Bartsch}},\ }\href {\doibase 10.1103/PhysRevLett.89.125701} {\bibfield
  {journal} {\bibinfo  {journal} {Phys. Rev. Lett.}\ }\textbf {\bibinfo
  {volume} {89}},\ \bibinfo {pages} {125701} (\bibinfo {year}
  {2002})}\BibitemShut {NoStop}%
\bibitem [{\citenamefont {Voigtmann}(2011)}]{Voigtmann:2011}%
  \BibitemOpen
  \bibfield  {author} {\bibinfo {author} {\bibfnamefont {{\relax
  Th}.}~\bibnamefont {Voigtmann}},\ }\href
  {http://stacks.iop.org/0295-5075/96/i=3/a=36006} {\bibfield  {journal}
  {\bibinfo  {journal} {Europhys. Lett.}\ }\textbf {\bibinfo {volume} {96}},\
  \bibinfo {pages} {36006} (\bibinfo {year} {2011})}\BibitemShut {NoStop}%
\bibitem [{\citenamefont {Zaccarelli}\ \emph {et~al.}(2005)\citenamefont
  {Zaccarelli}, \citenamefont {Mayer}, \citenamefont {Asteriadi}, \citenamefont
  {Likos}, \citenamefont {Sciortino}, \citenamefont {Roovers}, \citenamefont
  {Iatrou}, \citenamefont {Hadjichristidis}, \citenamefont {Tartaglia},
  \citenamefont {L\"owen},\ and\ \citenamefont
  {Vlassopoulos}}]{Zaccarelli:2005}%
  \BibitemOpen
  \bibfield  {author} {\bibinfo {author} {\bibfnamefont {E.}~\bibnamefont
  {Zaccarelli}}, \bibinfo {author} {\bibfnamefont {C.}~\bibnamefont {Mayer}},
  \bibinfo {author} {\bibfnamefont {A.}~\bibnamefont {Asteriadi}}, \bibinfo
  {author} {\bibfnamefont {C.~N.}\ \bibnamefont {Likos}}, \bibinfo {author}
  {\bibfnamefont {F.}~\bibnamefont {Sciortino}}, \bibinfo {author}
  {\bibfnamefont {J.}~\bibnamefont {Roovers}}, \bibinfo {author} {\bibfnamefont
  {H.}~\bibnamefont {Iatrou}}, \bibinfo {author} {\bibfnamefont
  {N.}~\bibnamefont {Hadjichristidis}}, \bibinfo {author} {\bibfnamefont
  {P.}~\bibnamefont {Tartaglia}}, \bibinfo {author} {\bibfnamefont
  {H.}~\bibnamefont {L\"owen}}, \ and\ \bibinfo {author} {\bibfnamefont
  {D.}~\bibnamefont {Vlassopoulos}},\ }\href {\doibase
  10.1103/PhysRevLett.95.268301} {\bibfield  {journal} {\bibinfo  {journal}
  {Phys. Rev. Lett.}\ }\textbf {\bibinfo {volume} {95}},\ \bibinfo {pages}
  {268301} (\bibinfo {year} {2005})}\BibitemShut {NoStop}%
\bibitem [{\citenamefont {Mayer}\ \emph {et~al.}(2008)\citenamefont {Mayer},
  \citenamefont {Zaccarelli}, \citenamefont {Stiakakis}, \citenamefont {Likos},
  \citenamefont {Sciortino}, \citenamefont {Munam}, \citenamefont {Gauthier},
  \citenamefont {Hadjichristidis}, \citenamefont {Iatrou}, \citenamefont
  {Tartaglia},\ and\ \citenamefont {Vlassopoulos}}]{Mayer:2008}%
  \BibitemOpen
  \bibfield  {author} {\bibinfo {author} {\bibfnamefont {C.}~\bibnamefont
  {Mayer}}, \bibinfo {author} {\bibfnamefont {E.}~\bibnamefont {Zaccarelli}},
  \bibinfo {author} {\bibfnamefont {E.}~\bibnamefont {Stiakakis}}, \bibinfo
  {author} {\bibfnamefont {C.~N.}\ \bibnamefont {Likos}}, \bibinfo {author}
  {\bibfnamefont {F.}~\bibnamefont {Sciortino}}, \bibinfo {author}
  {\bibfnamefont {A.}~\bibnamefont {Munam}}, \bibinfo {author} {\bibfnamefont
  {M.}~\bibnamefont {Gauthier}}, \bibinfo {author} {\bibfnamefont
  {N.}~\bibnamefont {Hadjichristidis}}, \bibinfo {author} {\bibfnamefont
  {H.}~\bibnamefont {Iatrou}}, \bibinfo {author} {\bibfnamefont
  {P.}~\bibnamefont {Tartaglia}}, \ and\ \bibinfo {author} {\bibfnamefont
  {D.}~\bibnamefont {Vlassopoulos}},\ }\href {\doibase 10.1038/nmat2286}
  {\bibfield  {journal} {\bibinfo  {journal} {Nature Materials}\ }\textbf
  {\bibinfo {volume} {7}},\ \bibinfo {pages} {780} (\bibinfo {year}
  {2008})}\BibitemShut {NoStop}%
\bibitem [{\citenamefont {Deutschl\"ander}\ \emph {et~al.}(2014)\citenamefont
  {Deutschl\"ander}, \citenamefont {Puertas}, \citenamefont {Maret},\ and\
  \citenamefont {Keim}}]{Deutschlaender:2014}%
  \BibitemOpen
  \bibfield  {author} {\bibinfo {author} {\bibfnamefont {S.}~\bibnamefont
  {Deutschl\"ander}}, \bibinfo {author} {\bibfnamefont {A.~M.}\ \bibnamefont
  {Puertas}}, \bibinfo {author} {\bibfnamefont {G.}~\bibnamefont {Maret}}, \
  and\ \bibinfo {author} {\bibfnamefont {P.}~\bibnamefont {Keim}},\ }\href
  {\doibase 10.1103/PhysRevLett.113.127801} {\bibfield  {journal} {\bibinfo
  {journal} {Phys. Rev. Lett.}\ }\textbf {\bibinfo {volume} {113}},\ \bibinfo
  {pages} {127801} (\bibinfo {year} {2014})}\BibitemShut {NoStop}%
\bibitem [{\citenamefont {Klix}\ \emph {et~al.}(2012)\citenamefont {Klix},
  \citenamefont {Ebert}, \citenamefont {Weysser}, \citenamefont {Fuchs},
  \citenamefont {Maret},\ and\ \citenamefont {Keim}}]{Klix:2012}%
  \BibitemOpen
  \bibfield  {author} {\bibinfo {author} {\bibfnamefont {C.~L.}\ \bibnamefont
  {Klix}}, \bibinfo {author} {\bibfnamefont {F.}~\bibnamefont {Ebert}},
  \bibinfo {author} {\bibfnamefont {F.}~\bibnamefont {Weysser}}, \bibinfo
  {author} {\bibfnamefont {M.}~\bibnamefont {Fuchs}}, \bibinfo {author}
  {\bibfnamefont {G.}~\bibnamefont {Maret}}, \ and\ \bibinfo {author}
  {\bibfnamefont {P.}~\bibnamefont {Keim}},\ }\href {\doibase
  10.1103/PhysRevLett.109.178301} {\bibfield  {journal} {\bibinfo  {journal}
  {Phys. Rev. Lett.}\ }\textbf {\bibinfo {volume} {109}},\ \bibinfo {pages}
  {178301} (\bibinfo {year} {2012})}\BibitemShut {NoStop}%
\bibitem [{\citenamefont {Mazoyer}\ \emph {et~al.}(2009)\citenamefont
  {Mazoyer}, \citenamefont {Ebert}, \citenamefont {Maret},\ and\ \citenamefont
  {Keim}}]{Mazoyer:2009}%
  \BibitemOpen
  \bibfield  {author} {\bibinfo {author} {\bibfnamefont {S.}~\bibnamefont
  {Mazoyer}}, \bibinfo {author} {\bibfnamefont {F.}~\bibnamefont {Ebert}},
  \bibinfo {author} {\bibfnamefont {G.}~\bibnamefont {Maret}}, \ and\ \bibinfo
  {author} {\bibfnamefont {P.}~\bibnamefont {Keim}},\ }\href
  {http://stacks.iop.org/0295-5075/88/i=6/a=66004} {\bibfield  {journal}
  {\bibinfo  {journal} {EPL (Europhysics Letters)}\ }\textbf {\bibinfo {volume}
  {88}},\ \bibinfo {pages} {66004} (\bibinfo {year} {2009})}\BibitemShut
  {NoStop}%
\end{thebibliography}

%

\end{document}